RWTH Aachen University
Software Engineering

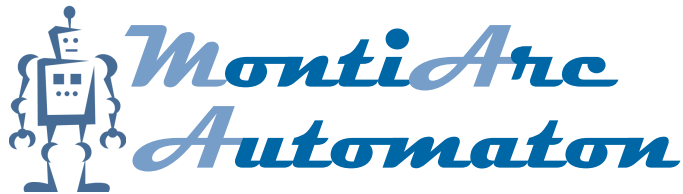

# Architecture and Behavior Modeling of Cyber-Physical Systems with MontiArcAutomaton

Jan Oliver Ringert
Bernhard Rumpe
Andreas Wortmann

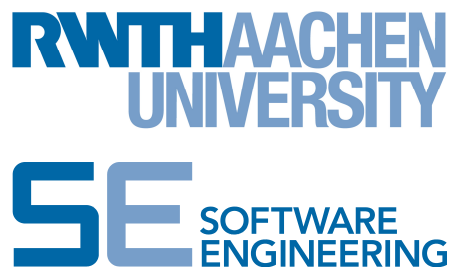

February 27, 2015

# Abstract

This book presents MontiArcAutomaton, a modeling language for architecture and behavior modeling of Cyber-Physical Systems as interactive Component & Connector models. MontiArcAutomaton extends the Architecture Description Language MontiArc with automata to describe component behavior.

The modeling language MontiArcAutomaton provides syntactical elements for defining automata with states, variables, and transitions inside MontiArc components. These syntactical elements and a basic set of well-formedness rules provide the syntax for a family of modeling languages for state-based behavior modeling in Component & Connector architectures. We present two concrete language profiles with additional well-formedness rules to model time-synchronous component behavior and untimed, event-driven behavior of components.

This book gives an overview of the MontiArcAutomaton language including examples, a language reference, and a context-free grammar for MontiArcAutomaton models. It also provides syntax definition, well-formedness rules, and semantics for two language profiles. We summarize projects and case studies applying MontiArcAutomaton.

MontiArcAutomaton is implemented using the DSL framework MontiCore. Available tools include a textual editor with syntax highlighting and code completion as well as a graphical editor and a powerful and extensible code generation framework for target languages including EMF, Java, Mona, and Python.

# Contents

# Chapter 1.

# Introduction

Cyber-Physical Systems (CPS) [Lee06] are networks of cooperating systems with both physical and digital input and output. Common applications for CPS are sensor-based systems, such as autonomous cars, smart grid, distributed robotics, and wireless sensor networks. The technological and social challenges arising from CPS [Lee08, BCG12] pervade many disciplines and implementation domains.

Efficient engineering of reliable and robust CPS requires new concepts, methods, and technologies from automation engineering [KRS12] to security [CAS08] to software engineering [Lee10]. A common concept to tame the increasing complexity of modern distributed software systems is the separation of concerns through modularization and decomposition into smaller parts. Component-based Software Engineering (CBSE) [McI68b] is a prominent, maturing, and successfully applied realization of this concept [HC01, BKM+05, NFBL10, SSL11]. Software components hide parts of the system's complexity behind well-defined, stable interfaces, which allows to develop and evolve system parts independently by respective experts. Yet, the integration of cyber modules and physical parts for non-trivial systems requires tremendous effort due to the "conceptual gap" [FR07] between problem domains (e.g., autonomous navigation) and implementation domains (e.g., software engineering).

Bridging the conceptual gap by handcrafting such software systems introduces "accidental complexities" [FR07] (such as dealing with specific API issues rather than solution concepts) which increase costs and difficulty of the software engineering process. Model-Driven Engineering (MDE) [Sel03, SVEH05] lifts models, rather than source code, to be primary development artifacts. These models describe different aspects of systems from various perspectives and at multiple levels of abstraction. Using sophisticated toolchains, such models are transformed into running systems. As models abstract from implementation details, MDE introduces less accidental complexities and thus reduces the conceptual gap between problem domains and implementation domain.

Many architectures of pure software systems as well as CPS with large software parts are modeled as Component and Connector (C&C) architectures [MT00, TMD09]. Components encapsulate a related subset of a system's functionality or data and define explicit interfaces to restrict access to these services (see [TMD09]). Connectors establish and regulate communication of components. Typically, connectors connect ports of components with compatible interfaces to allow interaction. Component-based development of software-intensive systems yields many benefits [McI68a, TMD09] as it facilitates reuse and enables physically as well as logically distributed development of software systems.

Modeling languages for software and CPS architectures are called Architecture Description Languages (ADLs). MontiArc [HRR12] is an ADL implemented as a textual Domain Specific Language (DSL) on top of the DSL framework MontiCore [GKR$^+$08, KRV08]. ADLs like MontiArc have demonstrated their usefulness in various domains to describe the structure of software systems.

We describe MontiArcAutomaton (MAA), an extension of the ADL MontiArc with an integrated specification mechanism to model component behavior. This is implemented in the modeling language MontiArcAutomaton (MAA), which extends MontiArc by embedding automata into components to model their behavior. The extension with syntactical elements makes MontiArcAutomaton a language family allowing the definition of different concrete language profiles. Syntax and semantics of MontiArcAutomaton are based on the I/O$^\omega$ automata paradigm [Rum96] and the Focus framework [BS01, RR11]. I/O$^\omega$ automata are automata, which allow reading from and sending messages to the ports of their encompassing component, thus partially reducing the need for behavior programming with general purpose programming languages such as Java or Python as demonstrated in [RRW13a, RRW13b].

It is important to note that there are multiple kinds of automata that can syntactically be expressed using MontiArcAutomaton. It is thus necessary to define language profiles and language profile specific semantics (in the sense of *meaning* [HR04b]). We present two language profiles in Chapter 4. One language profile supports time-synchronous component interaction while the other one provides automata for handling event-based communication.

Chapter 2 briefly introduces MontiArc [HRR12] before Chapter 3 illustrates MontiArcAutomaton by example. Chapter 4 presents two language profiles with syntactically specialized automata and corresponding automata semantics. The subsequent Chapter 5 is a reference for MontiArcAutomaton language elements. Chapter 6 lists context conditions required for the well-formedness of MontiArcAutomaton models. Chapter 7 describes several case studies with MontiArcAutomaton and different robotics platforms and Chapter 8 discusses the experiences with MontiArcAutomaton and directions for future work. Finally, Chapter 9 concludes this work.

# Chapter 2.

# The Architecture Description Language MontiArc

MontiArc [HRR12] is a modeling language for the description of C&C software architectures inspired by Focus [BS01] and C&C ADLs [TMD09]. Information is exchanged via connectors between the typed and directed ports of component interfaces. The component and connector concept allows composition of complex component hierarchies where components are either *atomic*, and perform execution of functionalities themselves, or are hierarchically *composed* of other components. Modeling component-based architectures offers benefits over traditional CBSE as models are platform independent, better comprehensible, and can be used to analyze properties of architectures based on an abstract model (cf. [TMD09, HRR12]). This section provides an overview of the ADL MontiArc by example of a simple robot.

The bumper bot robot comprises an ultrasonic sensor to detect the distance to the closest obstacle in front, a controller which reacts to obstacles, and two motors which propel two parallel tracks. After the robot is activated, it explores a room by driving straight forward until an obstacle is detected, then it backs up, rotates, and continues to drive forward again (cf. [RRW13a]).

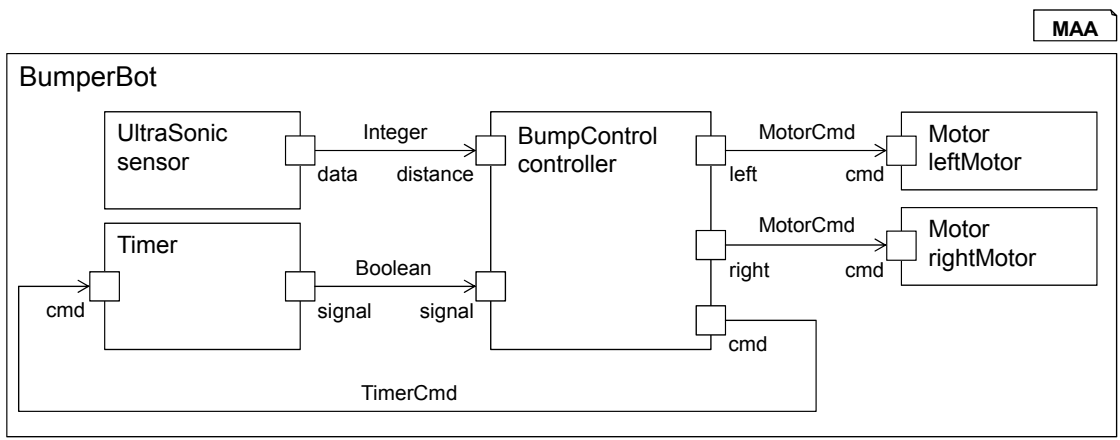

Figure 2.2.: Architecture of the composed component `BumperBot` for the bumper bot robot with its subcomponents.

Figure 2.2 depicts the software architecture for a bumper bot robot. The composed top-level component `BumperBot` consists of five subcomponents: `sensor` of component type `UltraSonic`, `controller` of component type `BumpControl`, `leftMotor` and `rightMotor` of component type `Motor`, and `timer` of component type `Timer`. The components exchange messages via the ports of their interfaces. Ports have a name, a type, and a direction. Their types are defined in UML/P class diagrams [Sch12] and restrict what messages they may send or receive – and thus also determine partners for possible connections. Connectors can only be established between ports of the same or of a compatible type. Here, component `sensor` sends distance measurements of type `Integer` via its outgoing port `data` to the controller's incoming port `distance` of the same type. Furthermore, only ports in the same scope can be connected directly, i.e., a port of `sensor` can neither be connected directly to a port outside the component `BumperBot`, nor directly to a subcomponent of `controller`. Figure 2.4 illustrates the enumeration types used by the component `controller` to send messages to the timer and the motors.

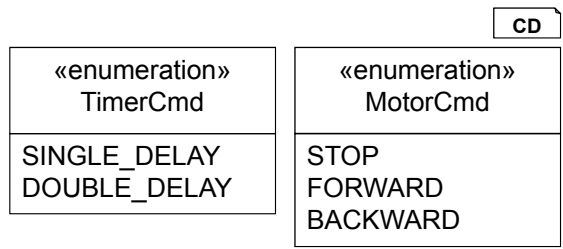

Figure 2.4.: Class diagram defining the two enumerations `TimerCmd` and `MotorCmd` used by the component `BumperBot` and its subcomponents.

After the `controller` has received a distance measurement and a timer signal it determines the course of action (i.e., either continues driving forward, backs up, or rotates) and sends according messages via its outgoing ports. Communication between MontiArc components is based on Focus [BS01], a framework for specifying and modeling distributed systems. Messages are passed asynchronously via typed unidirectional channels. The observable behavior on a channel is modeled as a finite or infinite stream of messages in the order of their transmission [HRR12].

The MontiArc code generation and simulation framework [wwwb] generates Java code and schedulers for time-synchronous and asynchronously timed communication [RR11]. In Section 4.1 we introduce a language profile of MontiArcAutomaton that implements time-synchronous communication of components.

The behavior of composed components emerges from the composition of the behaviors of their subcomponents. Interfaces of components do not reveal information about their behavior or possible composition, thus each of the subcomponents of `BumperBot` may be further composed. The distinction between the interface of a component and its behavior allows to introduce alternative behavior implementations. While the simulation framework of MontiArc only supports Java component behavior implementations, Monti-ArcAutomaton extends MontiArc to enable modeling of component behavior as I/O$^\omega$ automata [RRW13b]. The next chapter illustrates this extension.

## 2.1. Applications and Extensions of MontiArc

MontiArc has been extended in the context of modeling product variability [HRR$^+$11] using deltas [HRRS11]. Delta modeling is a bottom up technique starting with a small, but complete base variant. Features are added (that sometimes also modify the core). A set of applicable deltas configures a system variant. [HRR$^+$11, HRRS11] discuss the application of this technique to Delta-MontiArc. Deltas can not only describe spacial variability but also temporal variability which allows for using them for software product line evolution [HRRS12].

[GHK$^+$07] and [GHK$^+$08] provide means for modeling requirements on the structure of the logical architecture of interactive systems. The implementation of a specification language for crosscutting structural C&C views [MRR13, MRR14] is based on MontiArc. Prototype tools allow the verification of C&C models against C&C views [MRR14] and a synthesis prototype computes a satisfying C&C model for valid, invalid, alternative, and dependent C&C views, if one exists [MRR13].

A variant of MontiArc for cloud-based software architectures [NPR13] introduces cloud-specific language elements, e.g., replicating components and message groups. The extended framework supports code generation to Java and features solutions for typical cloud-computing challenges, such as event-based communication and serialization into databases.

# Chapter 3.

# MontiArcAutomaton Example and Overview

MontiArcAutomaton extends MontiArc by introducing automata to define component I/O behavior. Our automata are platform independent, i.e., they allow to model component behavior independently of target platforms and languages. By using different code generators, one can execute the same architecture and automata on different platforms.

This chapter illustrates the extensions of MontiArcAutomaton over MontiArc on the example of component `controller` of type `BumpControl`. This component is used to control the actions of the bumper bot robot introduced in Figure 2.2. In MontiArcAutomaton component behavior is modeled as an automaton. Figure 3.2 depicts this automaton embedded into the surrounding component.

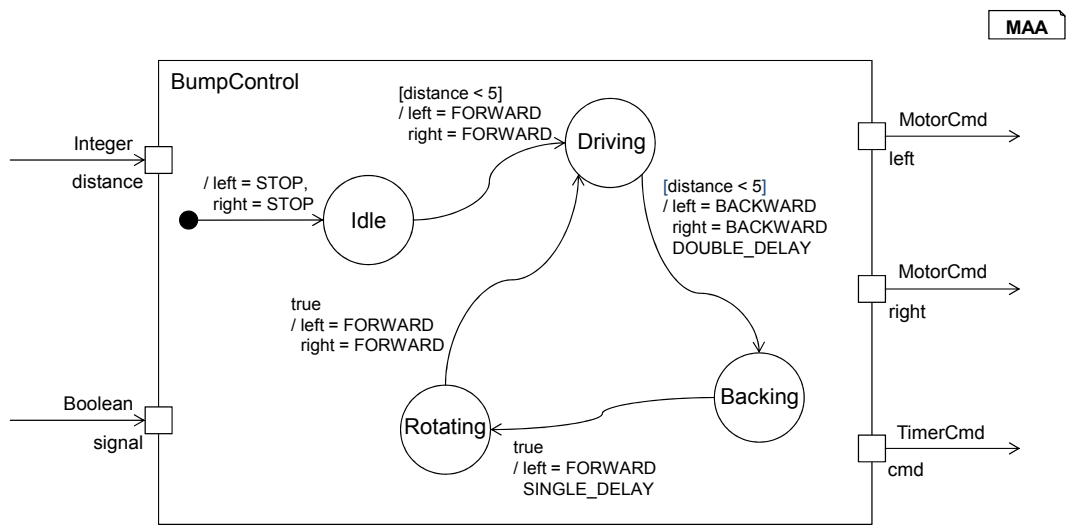

Figure 3.2.: The atomic component `BumpControl` with an embedded automaton.

The automaton consists of the four states `Idle`, `Driving`, `Backing`, and `Rotating` and five transitions. Generally, the automata of MontiArcAutomaton are finite in their number of states and support an arbitrary number of initial states. Each transition has a source state, a target state (which may be the same as the source state), and can be labeled with a guard, an input block, and an output block. Guards, input blocks, and output blocks of transitions are conceptually connected to the encompassing component: they either refer to ports of the component, which the automaton is embedded in, or its local variables in combination with appropriate messages and values for the corresponding port and variable types.

An input block states which input is necessary to activate the respective transition. Therefore, each input block may reference an arbitrary subset of input ports and variables to state which messages, according to the port and variables types, must be received. Output blocks define the messages sent out via the components output ports and variable assignments. This allows automata to model observable component behavior.

Here, the transition from `Backing` to `Rotating` denotes, that, if the messages received on incoming port `signal`[1] is `true`, the message `FORWARD` is send via the outgoing port `left` and the message `SINGLE_DELAY` is send via the outgoing port `cmd`[2]. Guards are conditions over the received input and the value of local variables that restrict when transitions may be executed. They can be formulated using the Object Constraint Language (OCL) variant of UML/P [Sch12] or as Java expressions. The automaton may only execute a transition if the guard condition holds. In this example, the transition form `Idle` to `Driving` may only occur, if the guard condition requiring that the value on incoming port `distance` is less than 5 is satisfied. Then the message `FORWARD` is send via the outgoing ports `left` and `right`.

MontiArcAutomaton is implemented as a textual Domain Specific Language (DSL) with the MontiCore [GKR+08, Vö11, Sch12, wwwc] language workbench. The formalisms for textual descriptions of hierarchically structured components are the same as in MontiArc. Therefore our examples focus on components models defining behavior. A textual representation of component `BumpControl` in Figure 3.2 is shown in Listing 3.3.

The organization of component definitions in files is similar to Java, as each compilation unit (model file) may contain only one parent component definition and these compilation units are organized in packages (Listing 3.3, l. 1). Packages correspond to the directory structure in the model path. Other compilation units (e.g., other components or types) can be used by importing their packages (l. 3). Component definitions start with the keyword `component` preceding the component name (l. 5). Typically, each MontiArc-Automaton component declares its interface at the beginning of the component body. An interface declaration starts with the keyword `port` and is followed by a set of ports which are labeled either as input ports (keyword `in`) or as output ports (keyword `out`) (ll. 7-12). Here each port is assigned a type and a name where the type is an unqualified reference to a previously imported data type.

The component `BumpControl` contains an automaton definition initiated by the keyword `automaton` (l. 14). The subsequent automaton body defines states and transitions in arbitrary order. States are declared by the keyword `state` followed by at least one state name (l. 15). Following the notation of [Rum96], initial states are defined explicitly using the keyword `initial` followed by a least one state name (l. 17).

---

[1]If the type of a value unambiguously matches a single port or variable, MontiArcAutomaton allows to omit the port's or variable's name in input blocks and output blocks.

[2]Again via type inference.

```
                                                    MontiArcAutomaton
 1  package bumperbot;
 2
 3  import bumperbot.types.*;
 4
 5  component BumpControl {
 6
 7    port
 8      in Integer distance,
 9      in Boolean signal,
10      out MotorCmd left,
11      out MotorCmd right,
12      out TimerCmd cmd;
13
14    automaton {
15      state Idle, Driving, Backing, Rotating;
16
17      initial Idle / {right = STOP, left = STOP};
18
19      Idle -> Driving [distance < 5]
20          / {left = STOP, right = STOP};
21
22      Driving -> Backing [distance < 5]
23          / {left = BACKWARD, right = BACKWARD, DOUBLE_DELAY};
24
25      Backing -> Rotating {true} / {left = FORWARD, SINGLE_DELAY};
26
27      Rotating -> Driving true / {left = FORWARD, right = FORWARD};
28    }
29  }
```

Listing 3.3: The component `BumpControl` in textual syntax

Transitions are not declared by a designated keyword but are instead defined by their
unique syntax. A transition declaration is initiated by a source state name followed by
a target state name (ll. 19-27). Further elements of a transition definition are source
state, target state, guard, input block, and output block. For transitions looping from
a state to itself, denoting the target state is optional. Subsequently, a guard may be
defined by providing an OCL or a Java expression in square brackets (ll. 19 and 22). For
each transition an optional input block may follow (ll. 25 and 27). This block specifies
messages read on incoming ports and values of local variables, which enable the transition
(e.g., the message `true` in l. 25). If the type of a value uniquely identifies the port (or
variable), the name of the port (or variable) can be omitted. For instance, the value
`true` of type `Boolean` used as input on the transition in from `Backing` to `Rotating`
(l. 25) can only be read from port `signal` as this is the only port of the same type.

The curly brackets for input blocks and output blocks can be omitted (compare ll. 25 and 27) and serve merely a structuring purpose. Finally, each transition may define an output block, which specifies messages or sequences of messages sent on output ports and values assigned to variables. Similar to input blocks, curly brackets and port names are optional as well. Guards, input blocks, and output blocks are optional and thus may be left out.

## 3.1. Related Modeling Languages

Similar approaches to integrated modeling of architecture structure with component behavior for CPS are the Architecture Analysis & Design Language [FG12] (AADL), AutoFOCUS [HF11], Simulink [Tya12], SysML [Wei06, FMS11], and UML composite structure diagrams with statecharts [OMG10].

AADL is a modeling language for hardware and software of embedded systems and as such also features constructs to model hardware components, while MontiArcAutomaton focuses on modeling logical software components. In AADL, components can be of component type thread, which may define sequences of *subprogram* calls. A subprogram comprises a component-like interfaces and "represents callable unit of sequentially executable code" [FG12]. AADL can also be extended with behavior modeling languages through sublanguage conforming to the behavior annex [BFBFR07], which lacks integrated semantics with the surrounding architecture [YHMP09].

AutoFOCUS is a modeling tool and C&C ADL for the development of distributed embedded systems which is also based on the formal semantics of FOCUS [BS01, RR11]. AutoFOCUS supports timesynchronous streams with strongly causal and weakly causal component behavior. Behavior is modeled as state transition diagrams similar to MAA automata. In contrast to MontiArcAutomaton, AutoFOCUS lacks a distinction between component types and their instantiations, which hampers reuse of components.

MathWorks Simulink features a block diagram language enabling the description of components and connectors. Stateflow [wwwa] extends blocks with state transition diagrams. The semantics of Stateflow is not completely defined and has been formalized in different ways [HR04a, MC12]. In contrast to the automata of MontiArcAutomaton, the automata of stateflow do neither support underspecification nor refinement.

SysML is a graphical modeling language family for the development of software systems based on a subset of extended UML [OMG10] languages. SysML features languages to describe requirements, structure and behavior of systems. System structure is captured in block definition diagrams, internal block diagrams, and package diagrams. Internal block diagrams feature components (called "parts"), connectors, and ports and thus are similar to MontiArc models. System behavior is captured with activity diagrams, sequence diagrams, state machine diagrams and use case diagrams. As MontiArcAutomaton automata can be considered a language profile of UML statecharts, SysML enables to express architectures similar to MontiArcAutomaton architectures. In contrast to SysML, the semantics of a MontiArcAutomaton architecture is well-defined and grounded in the FOCUS framework, while the semantics of SysML models is grounded in the code genera-tor employed. The same holds for the combination of UML composite structure diagrams with statecharts.

# Chapter 4.

# A MontiArcAutomaton Language Profiles and Semantics

MontiArcAutomaton extends MontiArc [HRR12] with syntax for automata inside component definitions. The extended syntax offers support for specifying states, variables, and transitions. Transitions feature messages on input ports, message sequences on output ports, guards, and variables. A formal definition of the concrete and abstract syntax of MontiArcAutomaton is given as MontiCore grammar in the appendix (see Appendix A for a simplified human-readable version of the grammar and Appendix B for the detailed MontiCore grammar). We provide examples of all syntactical elements in Chapter 5. Chapter 6 lists context conditions that well-formed MontiArcAutomaton models need to satisfy.

A modeling language definition consists not only of the concrete and abstract syntax of the language but also of its semantics (in the sense of *meaning* [HR04b]). For MontiArc-Automaton we chose the FOCUS calculus [BS01, RR11] of streams and stream processing functions as the semantic domain. This semantic domain allows to represent the interaction behavior of various kinds of systems described by automata [Rum96, BR07, BCR07, BCGR09]. We sketch the semantics of specific language profiles of MontiArcAutomaton by giving examples for the execution of automata, i.e., the input streams they consume and the output streams they produce.

One may define multiple language profiles that are each suited for different modeling purposes. A general discussion of syntactic and semantic variability in modeling language definitions is presented in [CGR09]. The modeling language MontiArcAutomaton forms a superset of syntactical elements to express several kinds of automata. Its syntax can be restricted for specific language profiles by additional well-formedness rules as shown by the examples in Section 6.5 and Section 6.6. In addition the syntax of MontiArc-Automaton can be extended or the semantics specialized using stereotypes at various places.

In the following, we present two language profiles and sketch their semantics for time-synchronous and event-driven communication [RR11]. One profile restricts the structure of automata to model time-synchronous automata. The other is a variant of event-driven automata. We sketch the semantics of the automata for these language profiles by giving examples of automata executions.

# 4.1. Language Profile for Time-Synchronous Communication

We present the profile MAA$_{ts}$ of the MontiArcAutomaton language to model interactive components using time-synchronous communication. The key idea of time-synchronous communication is an execution of the system in discrete global time cycles. Each cycle corresponds to the execution of one transition of each MAA$_{ts}$ automaton or an idle cycle if none of its transitions is enabled. During a cycle a component reads its inputs and sends outputs that are then available as input to the communication partners in the next cycle. This MontiArcAutomaton language profile restricts the output in every time cycle to at most one message per port. Each transition may read all variables and all messages currently available on input ports, write up to one message to each output port, and assign new variable values.

In typical CPS we observe various communication behavior. Components measuring or aggregating sensor data may send messages with the most recent data in every time cycle. Components on a higher level of abstraction may send command messages in one time cycle but then wait for an event, e.g., receiving a response or observing the change of sensor data in future time cycles, before issuing further commands. We thus include handling of the absence of messages in the syntax and semantics of time-synchronous automata. The MontiArcAutomaton language profile adds a special symbol `--` (see productions `NoData` and `OptVal` in the MontiArcAutomaton grammar in Listing B.1) to allow modelers to specify the expected or forced absence of messages.

The structure of MAA$_{ts}$ automata is given in Definition 4.1. The well-formedness rules for the MontiArcAutomaton syntax to conform to the MAA$_{ts}$ language profile are described in Chapter 6. Specific well-formedness rules for the MAA$_{ts}$ profile are defined in Section 6.5.

**Definition 4.1** (Time-synchronous MontiArcAutomaton (MAA$_{ts}$)). *A time-synchronous MAA is a syntactically restricted automaton where*

- *the output is at most one message per port and*

- *the special message `--` may occur on all input and output ports.*

△

MAA$_{ts}$ automata are executed in cycles where in each time cycle one enabled transition is executed, if one exists. Transitions are enabled if the input messages and variable values specified in the input block match the messages read on the input ports and the values of the local variables in the current time cycle. We interpret the absence of declared messages and variable values as underspecification by the modeler, that is, the message on a port and value of a variable omitted in the input block is not relevant for enabling the transition, i.e., all possible messages and values are allowed on the omitted port or variable. We interpret the absence of outputs on a transition as not sending a message on that port (denoted as `--` in the time-synchronous semantics). The absence of assignments to variables on a transition is interpreted as preserving variable values.

The messages on all input ports are available only in the cycle after they have been sent. In case a message is not read on a port it is not buffered and then is lost in the next cycle. If no transition is enabled in one execution cycle, the automaton does not execute any transition and therefore does not produce any message in the current time cycle (again denoted as `--`). This behavior can also be interpreted as the execution of a self-loop, with output `--` on every port, that is enabled if no existing transition is enabled (so called *idle completion* [Rin14, Def. 6.25]). Again, in the next cycle the previous inputs are no longer available on the port.

As an example for an automaton modeled according to the MAA$_{ts}$ profile, consider multiple vehicles driving in a convoy. The controller of a vehicle contains a component that decides the speed of driving forward based on the information whether the vehicle is in its lane and based on the distance to the leading vehicle. Part of this component is sketched in the MAA$_{ts}$ automaton shown in Listing 4.2. Component `FollowThe-LeaderOnline` controls the speed of going forward by sending commands to the motor. If the automaton is in the state `Following`, if the vehicle is in its lane, and if the distance measured has the value `TOO_FAR`, the port `cmd` sends the command `FAST_FORWARD`.

```
                                                          MontiArcAutomaton-TS
1  package robot;
2
3  component FollowTheLeaderOnline {
4
5    port
6      in Boolean inLane,
7      in Distance dist,
8      out MotorCmd cmd;
9
10   automaton {
11     state Following, Finding, Waiting;
12     initial Following / SLOW_FORWARD;
13
14     Following {inLane = true, dist = TOO_FAR} / FAST_FORWARD;
15     Following {inLane = true, dist = TOO_CLOSE} / SLOW_FORWARD;
16     Following -> Finding false / TURN;
17
18     // ... more transitions  ...
19   }
20 }
```

Listing 4.2: The MAA$_{ts}$ automaton of component `FollowTheLeaderOnline` to follow a leading object while staying in lane

Table 4.3 shows the reaction of the component to inputs given on the ports `inLane` and `dist`. Each column of the table contains the input and output messages of one cycle of the synchronous system execution. As an example, the initial output message `SLOW_-FORWARD` of the automaton (Listing 4.2, l. 12) is sent in the first cycle (see Table 4.3).

| cycle | **in** `inLane` | **in** `dist` | **out** `cmd` |
|:-----:|:---------------:|:-------------:|:--------------|
| 1 | `true` | `--` | `SLOW_FORWARD` |
| 2 | `true` | `--` | `--` |
| 3 | `true` | `TOO_FAR` | `--` |
| 4 | `true` | `TOO_FAR` | `FAST_FORWARD` |
| 5 | `true` | `--` | `FAST_FORWARD` |
| 6 | `false` | `--` | `--` |
| 7 | `false` | `TOO_FAR` | `TURN` |
| 8 | `true` | `TOO_FAR` | `--` |
| 9 | | | `SLOW_FORWARD` |

Table 4.3.: Inputs and outputs of component `FollowTheLeaderOnline` at nine execution cycles

Consider the second cycle where the message `true` is received on the port `inLane` and no message is received on the incoming port `dist` (denoted by `--`). This input combination does not trigger a transition in the current state `Following` and nothing is sent on port `cmd` in cycle $t+1$ (denoted by `--`). The transition in line 14 of Listing 4.2 is triggered by the input in cycle 3 and the component sends the message `FAST_FORWARD` in cycle 4. The same input pattern is received in cycle 4 and the message `FAST_FORWARD` is repeated in cycle 5.

Our semantics of the MAA$_{ts}$ language profile has strongly causal component behavior. Strong causality requires that a possible reaction to an input at cycle $t$ may only occur in cycle $t+1$ or later [BS01, RR11]. For MAA$_{ts}$ the reaction to an input at time $t$ (as defined by a transition) happens at time $t+1$.

We have implemented MontiCore code generators for the MAA$_{ts}$ language profile. These generators generate executable Java and Python code, analyzable Mona predicates, and EMF models as presented in [RRW13b, RRW13c]. The code generator from MontiArcAutomaton models to the Mona specification language enables, e.g., refinement checking between different MontiArcAutomaton components [Kir11]. We have evaluated our Java code generator in a one-semester student project about model-based robotic system development [RRW13c].

## 4.2. A Language Profile for Event-Driven Automata

A different language profile of MontiArcAutomaton is the MAA$_{ed}$ profile for event-driven automata. The execution of transitions in event-driven automata is triggered by receiving an event on a port of the component. The component can then produce a finite number of events to emit on its output ports. The structure of event-driven automata is defined in Definition 4.4. It is important to note that each transition can only be triggered by a single event. Thus it is not possible to read multiple events on a transition or read multiple events in guards of transitions.

The well-formedness rules for the MontiArcAutomaton syntax to conform to the MAA$_{ed}$ language profile are described in Chapter 6. Specific well-formedness rules for the MAA$_{ed}$ profile are defined in Section 6.6.

**Definition 4.4** (Event-driven MontiArcAutomaton (MAA$_{ed}$)). *An event-driven MAA is a syntactically restricted automaton where*

- *on every transition the input and guard are restricted to read exactly one input message on exactly one port and*

- *the symbol −− must not be used in the input or output on a transition.*

$\triangle$

As an example, consider the robotic arm shown in Figure 4.6 and the component `ToastArmController` shown in Listing 4.7. The MAA$_{ed}$ automaton inside component `ToastArmController` has a transition from the state `Idle` activated by event `PICK_-UP_TOAST` received on port `req` (see ll. 15-16). The event is handled by sending the sequence of commands `[MOVE_UP, TURN_RIGHT, OPEN, MOVE_DOWN, CLOSE]` on the output port `armCmd`.

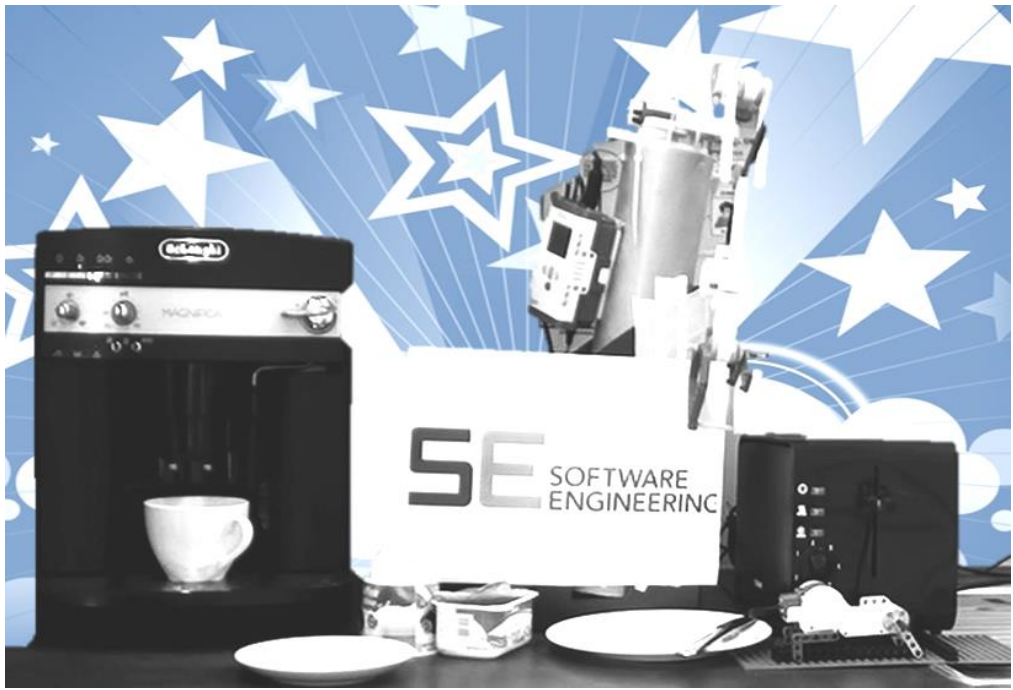

Figure 4.6.: A robotic arm capable of picking up a toast

Please note that a definition of the semantics of MAA$_{ed}$ automata based on stream processing functions requires a model of timed streams (see [RR11] for different kinds of

```
                                                          MontiArcAutomaton-ED
1  package robot;
2
3  component ToastArmController {
4
5    port
6      in Request req,
7      in Boolean reset,
8      out ArmControlCommand armCmd,
9      out LightCommand lightCmd;
10
11   automaton {
12     state Idle, GotToast;
13     initial Idle;
14
15     Idle -> GotToast PICK_UP_TOAST /
16        [MOVE_UP, TURN_RIGHT, OPEN, MOVE_DOWN, CLOSE], FLASH;
17     GotToast -> Idle DROP_TOAST /
18        [TURN_LEFT, MOVE_DOWN, OPEN], OFF;
19 }
```

Listing 4.7: The MAA$_{ed}$ automaton of a robotic arm controller for picking up and
        dropping toast

streams in FOCUS and see [Rum96] for semantics definitions of automata using stream
processing functions). Timed streams encode a model of time and thus make it possible
to relate the occurrence of events on different input and output ports of the component.

The language profile for MAA$_{ed}$ automata has been investigated for modeling and code
generation in a student project with an application to robotic systems [Mar12].

# Chapter 5.

# MontiArcAutomaton Language Reference

The language MontiArcAutomaton extends MontiArc with syntactical elements of I/O$^\omega$ automata [Rum96] and variables. This chapter introduces these language elements and their concrete syntax on the basis of small examples. The complete grammar of MontiArcAutomaton is available in Appendix A and Appendix B. Component definitions in MontiArcAutomaton can contain automata and variables as top level elements in addition to the language elements known from MontiArc [HRR12].

This chapter gives an overview over the syntactical elements of MontiArcAutomaton not contained in MontiArc. Chapter 6 introduces well-formedness rules common to all kinds of automata that can be expressed using the syntax of MontiArcAutomaton. Section 6.5 and Section 6.6 present well-formedness rules specific to the language profile for time-synchronous communication defined in Section 4.1 and specific to the profile for modeling event-driven automata defined in Section 4.2.

## 5.1. Automaton Declarations

An automaton declaration has to be contained in a component definition and starts with the keyword `automaton` followed by an optional name, and the body of the automaton. Optionally, stereotypes can be added to the declaration of the automaton before `automaton`.

Listing 5.1 shows the definition of component `IntegerBuffer` with an embedded automaton of name `BufferAutomaton` (l. 8).

## 5.2. Variables

Automata may reference variables in guards, input blocks, and output blocks. These variables are local to MontiArcAutomaton components and thus declared in the component's scope. A variable declaration consists of a type name, the variable's name and an optional initial value assignment. The types of variables are either defined in Java or UML/P class diagrams and need to be imported by the containing component.

Listing 5.2 depicts the declaration of a variable `buffer` of type `Integer` (l. 8).

<div style="border: 1px solid">MontiArcAutomaton</div>

```
 1  component IntegerBuffer1 {
 2
 3    port
 4      in Integer value,
 5      in Boolean saveValue,
 6      out Integer bufferedValue;
 7
 8    automaton BufferAutomaton {
 9      //...
10    }
11  }
```

Listing 5.1: Definition of an automaton with name `BufferAutomaton` inside component `IntegerBuffer`

<div style="border: 1px solid">MontiArcAutomaton</div>

```
 1  component IntegerBuffer2 {
 2
 3    port
 4      in Integer input,
 5      in Boolean saveValue,
 6      out Integer output;
 7
 8    Integer buffer;
 9
10    automaton BufferAutomaton {
11      //...
12    }
13  }
```

Listing 5.2: Variable declarations inside a MontiArcAutomaton component

## 5.3. Values and Sequences

Values assigned to variables and communicated via ports of the component use the production `Value` from the MontiArc grammar [HRR12] which provides values and literals for common types, e.g. `true` and `false` for type `Boolean`. Listing 5.3 illustrates the use of values: l. 13 defines a transition from state `S` to itself, which is enabled if a message of value `true` is received via the port `saveValue`. In this case, the input is saved to variable `buffer` and the value `0` is emitted via port `output`. If the value `false` is received via the port `saveValue`, the value stored in variable `buffer` is send via port `output`. As mentioned above, the port names can be omitted if the value types identify the intended port unambiguously (cf. inputs in ll. 13 and 14). As both transitions define a loop from state `S` to itself, explicitly defining the target state (i.e., `-> S`) could have been omitted.

```
                                                    MontiArcAutomaton
1  component IntegerBuffer3 {
2
3    port
4      in Integer input,
5      in Boolean saveValue,
6      out Integer output;
7
8    Integer buffer;
9
10   automaton BufferAutomaton {
11     state S;
12
13     S -> S true /  {output = 0, buffer = input };
14     S -> S false / {output = buffer};
15   }
16 }
```

Listing 5.3: Usage of literal values `true` and `0`

Some language profiles allow the sending of sequences of messages on output ports. A sequence is created from a list of values using the operator `[...]` with a comma-separated list of messages inside. The construction of message sequences is illustrated in Listing 5.4. Line 12 denotes a transition from state `S` to `T`, which reads both input ports and sends a sequence of the two values - not their concatenation - via its output port. Please note, that it is not necessary to specify the output port name `output`, as the sequence `[a, b]` consisting of two elements can only be send via port `output` of type `Integer`.

Finally, MontiArcAutomaton extends the `Value` production of the MontiCore grammar with the literal value `NoData` denoted by `--`. This (pseudo) value can be used in combination with a timed streams semantics [RR11] to denote the case of *no message available*. An example is given in the language profile MAA$_{ts}$ introduced in Section 4.1.

## 5.4. State Declarations

A state declaration introduces one or more states. An automaton can have multiple state declarations each starting with the keyword `state` followed by a list of states. A state has a name and can have an optional list of stereotypes. Stereotypes are provided as means to extend the language for generator developers and language profile developers. The interpretation of stereotypes is thus left to specific applications. Multiple states inside a state declaration are separated by a comma.

Listing 5.5 shows a single state declaration (l. 11) declaring two states with names `S` and `T`. The state `T` has the stereotype «error» which can be interpreted by applications.

MontiArcAutomaton

```
1  component Echo1 {
2    port
3      in Integer a,
4      in Integer b,
5      in Boolean speak,
6      out Integer output;
7
8    automaton {
9      state S, T;
10     initial S;
11
12     S -> T true / [a, b];
13   }
14 }
```

Listing 5.4: Sequences of values sent on output ports

MontiArcAutomaton

```
1  component IntegerBuffer4 {
2
3    port
4      in Integer input,
5      in String cmd,
6      out Integer output;
7
8    Integer buffer;
9
10   automaton BufferAutomaton {
11     state S, <<error>> T;
12
13     S -> S "SAVE" /  {output = 0, buffer = input};
14     S -> S "SEND" / output = buffer;
15     S -> T [cmd != "SAVE" && cmd != "SEND"];
16   }
17 }
```

Listing 5.5: Definition of states S, and T with state T having the stereotype «error»

## 5.4.1. Initial States and Initial Outputs

An automaton can declare multiple initial states using the keyword `initial` followed by at least one state name. Each of these state names can be assigned an initial output in terms of an output block.

Listing 5.6 introduces the state S (l. 11). State S is declared initial afterwards (l. 12) and defined to initially assigning the value 0 to the variable `buffer`. Please note, that

```
                                                    MontiArcAutomaton
 1  component IntegerBuffer5 {
 2
 3    port
 4      in Integer input,
 5      in Boolean saveValue,
 6      out Integer output;
 7
 8    Integer buffer;
 9
10    automaton BufferAutomaton {
11      state S;
12      initial S / buffer = 0;
13
14      S -> S true /  {output = 0, buffer = input};
15      S -> S false / {output = buffer};
16    }
17  }
```

Listing 5.6: Definition of an initial state `S` with initial output `0`

the curly brackets around assignments are optional again.

## 5.5. Transitions

Transitions are defined originating from a source state towards a target state. Omitting the target state defines a transition looping from the source state to itself. A transition further may have an optional guard, an optional input block, and an optional output block. See Listings 5.3 to 5.6 show various forms of syntactically valid transitions.

Transitions follow the pattern illustrated in Listing 3.3: a transition from state `S0` to state `S1` with guard `Guard`, inputs `Inputs`, and outputs `Outputs` generally has the form

                  S0 -> S1 [Guard] {Inputs} / {Outputs};

where

- `Guard` is an expression of the embedded guard language (currently available are OCL and Java).

- `Inputs` is a set of pairs of port names and variable names and their expected values that trigger the transition of the form `x = val`. If `x` is the name of a port, `val` must be a *message* consisting of a single *value*, i.e., no sequences of values. If `x` is the name of a variable, `val` must again be a single value. Alternatives for the values read on ports and variables may be written using the disjunction operator `|` as in the example `x = val1 | val2 | val3` to allow underspecification. Input

ports and variables may reference other input ports and variables as described in Section 5.5.2.

- `Outputs` is a set of pairs of port resp. variable names and assigned values that result from the execution of the transition and have the form `x = val`. If `x` is the name of a port, `val` may be a single *message* (e.g., `42`) or a sequence of messages (e.g., `[3, 14]`). If `x` is the name of a variable, `val` must be a single value. Non-deterministic alternatives for the values written on ports and assigned to variables may be written using the disjunction operator `|` as in the example `x = val1 | [val2, val3] | []`. Output ports and variable assignments may reference input ports and variables as described in Section 5.5.3.

The curly brackets can be omitted, such that `S0 -> S1 [Guard] Inputs / Outputs;` denotes the same transition as above. If the transition describes a loop from a state `S0` to itself, the target state may also be omitted, which results to the form `S0 [Guard] {Inputs} / {Outputs}`. Again, the curly brackets may be omitted — thus `S0 [Guard] Inputs / Outputs;` denotes the same transition. As input, guard, and output are also optional, the minimal transition has the form `S0`, which denotes an unconditional loop from S0 to itself that reads, emits, and assigns nothing. Whether this transition is allowed and what its semantics are is dependent on the language profile and semantics chosen.

### 5.5.1. Guard Expressions over Ports and Variables

Guards on transitions are surrounded by square brackets `[...]`. MontiArcAutomaton does not define its own expression language for guards. Current options are to use Java expressions of type Boolean (guard kind starting with the keyword `java:`) or OCL/P [Sch12] expressions (guard kind starting with the keyword `ocl:`). As OCL/P is the default guard expression language, specification of guard kind `ocl` can be omitted. Guards may refer to input ports and variables.

The optional OCL/P guard `[input <= 9]` in line 14 of Listing 5.7 expresses that its transition can only be activated if variable `input` has a value less or equal than 9. The second transition explicitly specifies to use a Java guard (l. 15) which holds, if the value of variable `input` is greater than 9.

### 5.5.2. Input on Ports and Current Variable Values

The optional input block of a transition contains, if present, at least one port or variable valuation. A valuation of a port specifies a value expected as input on this port. A valuation of a local variable specifies the value it shall have to trigger the transition. Multiple valuations of ports or variables are separated by commas. In case the port or variable name is uniquely determined by the type of the message or value, the name is optional in the input block. Alternatives for the values read on a port or variable with the name x may be written using the disjunction operator `|` as in the example `x = val1 | val2 | val3`.

```
                                                    ┌──────────────────────┐
                                                    │ MontiArcAutomaton    │
 ┌──────────────────────────────────────────────────┴──────────────────────┐
 1│component SmallNumbersBuffer {
 2│
 3│  port
 4│    in Integer input,
 5│    in Boolean saveValue,
 6│    out Integer output;
 7│
 8│  Integer buffer;
 9│
10│  automaton BufferAutomaton {
11│    state S;
12│    initial S / buffer = 0;
13│
14│    S [input <= 9] true / {output = 0, buffer = input};
15│    S [java: input >  9] true / {output = 0};
16│    S false / {output = buffer};
17│  }
18│}
```

Listing 5.7: Two guard expressions using the embedded OCL/P and Java respectively

```
                                                    ┌──────────────────────┐
                                                    │ MontiArcAutomaton    │
 ┌──────────────────────────────────────────────────┴──────────────────────┐
 1│component ZeroBuffer {
 2│
 3│  port
 4│    in Integer input,
 5│    in Boolean safe,
 6│    out Integer output;
 7│
 8│  Integer buffer = -1;
 9│
10│  automaton BufferAutomaton {
11│    state S;
12│    initial S;
13│
14│    S -> S true, input = 0;
15│    S -> S [input != 0] true;
16│    S -> S input = 0 | 1, false;
17│  }
18│}
```

Listing 5.8: A component with three transitions and different valuations

Listing 5.8 shows a component containing an automaton with three transitions over two input ports of types `Integer` and `Boolean` and a variable of type `Integer` as well. Using the value `true` in an input block unambiguously identifies the port `safe` of type

`Boolean` as the intended source, thus stating the the port's name can be omitted. Using numbers can either refer to the input port `input` or the variable `buffer`, therefore the name of the intended source has to be specified.

**Values in Input Blocks**

The right side of a port reference in the input is a disjunction over expected messages. A message can be a literal value conforming to the type of the port, the name of a variable, or the name of another input port. The right side of a variable reference can either be a literal value, or the name of another port or variable. Disjunctions of these are also allowed for references to variables.

### 5.5.3. Output on Ports

The optional output block of a transition starts with `/`. If the block is present, it has to contain at least one message as output to a port or an assignment of a value to a variable. The output sent via a port can either be a literal value, a variable name, the name of an input port, or a concatenation of any of these. The type of the assigned values has to correspond to the type of the port or variable. The list operator `[...]` concatenates messages to a list of messages, e.g., as list of String values as in Listing 5.9. Non-deterministic alternatives of values can be separated by the disjunction operator `|` inside an assignment. Multiple assignments are separated by commas.

```MontiArcAutomaton
component IntegerDuplicator {
  port
    in String input,
    in Boolean speak,
    out String output;

  automaton {
    state S;
    initial S;

    S false / --;
    S true / [input, input];
  }
}
```

Listing 5.9: Transitions producing two output messages of type `String`

Listing 5.9 shows a component with an automaton that produces two `Integer` output messages. The first transition (l. 11) is enabled if the message `false` was received on incoming port `speak` and emits the message `--` via outgoing port `output`. The second transition (l. 12) is enabled if the message `true` was received and sends a list of two `Integer` messages via outgoing port `output`. The output thus consists of two separate messages. As `output` is the only outgoing port that this sequence can be assigned to, it is not necessary to include the name of the port `output` in the output block.

### 5.5.4. Variable Value Assignments

Value assignments to variables may also appear in the output blocks of transitions. Same as with sending via ports, we use = after a variable name to denote an assignment of a value to the variable. The value assigned to a variable can either be a literal value, a variable name, or the name of an input port representing the value of a message read on that port. Non-deterministic alternatives of values can be separated by the disjunction operator | inside an assignment. Currently variable definitions (see Section 5.1) and the output blocks of transitions are the only places where values can be assigned to variables.

```
                                                    MontiArcAutomaton
1  component IntegerBuffer6 {
2
3    port
4      in Integer input,
5      in Boolean saveValue,
6      out Integer output;
7
8    Integer buffer;
9
10   automaton BufferAutomaton {
11     state S;
12
13     S -> S true /  {buffer = input, output = 0};
14     S -> S false / {buffer = 0,     output = buffer};
15   }
16 }
```

Listing 5.10: Setting the variable `buffer` on transitions

Listing 5.10 illustrates how the variable `buffer` of type `Integer` is set on two transitions (ll. 13 and 14). Using an `Integer` value in an output block might refer either to the outgoing variable `output` or the variable `buffer`, thus it is necessary to specify the port or variable names explicitly. Again, the curly brackets (ll. 13 and 14) are optional.

## 5.6. Generic Types

Generic component types are a mechanism to facilitate reuse of components. Component types can have generic type parameters which define the type of ports or variables. An example for a generic component type is the component type `Arbiter<T>` shown in Listing 5.11. Instantiations of the component `Arbiter<T>` need to provide a concrete type or another type variable for the parameter `T` of the component. The component

`Arbiter` uses the generic type `T` for the definition of its incoming and outgoing ports (see Listing 5.11, ll. 4-6).

```MontiArcAutomaton
component Arbiter<T> {
  port
    in Boolean mode,
    in T in1,
    in T in2,
    out T res;

  automaton {
    state S;
    initial S;
    S mode = true  / in1;
    S mode = false / in2;
  }
}
```

Listing 5.11: The generic `Arbiter` component selects and forwards one of two inputs based on the messages received on port `mode`

The use of generic component types and type variables is supported by MontiArc-Automaton. Type variables may be used for the types of ports and variables inside MontiArcAutomaton component definitions. In the example shown in Listing 5.11 the transition in line 11 forwards the message read on the input port `in1` to the output port `res` of the same type. Similarly, the transition in line 12 forwards the message read on port `in2` if the message read on port `mode` is `false`.

Additional well-formedness rules apply when using generic types, e.g., port names have to be given explicitly and are not derived from the type of the port. See [HRR12] for more details.

# Chapter 6.

# MontiArcAutomaton Context Conditions

The modeling language MontiArcAutomaton is defined by a context-free MontiCore grammar [GKR+08, KRV08]. Context-free grammars lack expressiveness to define various necessary properties (e.g., variables being defined twice in same scope) to make a model well-defined. MontiCore provides a powerful framework to describe these *context conditions* (CoCos) [Vö11]. The context conditions framework and the MontiCore symbol table framework [Vö11] allow, e.g., to check whether the value assigned to a port or variable has a compatible type.

We have implemented several context condition checks to assure the well-formedness of MontiArcAutomaton models. Further context conditions may be added depending on a chosen language profile and automata semantics (see Chapter 4) or to restrict models for the code generation to specific target languages. The context conditions of MontiArcAutomaton are divided into four groups regarding the nature of the checks. Our categorization of context conditions differentiates between requirements such as uniqueness of names (Section 6.1), conventions (Section 6.2), referential integrity (Section 6.3), and type correctness (Section 6.4). While these categories cover basic rules for the automaton language, additional context conditions for specified language profiles may be added (Section 6.5 and Section 6.6).

The extensions of MontiArcAutomaton which add automata and variables to the language require new context conditions while the existing MontiArc context conditions are inherited. A detailed list of MontiArc context conditions together with examples is given in the MontiArc technical report [HRR12]. In the following we describe the context conditions of MontiArcAutomaton by stating each condition as a rule and giving examples of context condition violations. At first, context conditions common to all possible language profiles are presented and afterwards (Section 6.5 and 6.6) language profile specific context conditions are presented.

## 6.1. Uniqueness Conditions

In order to create correct models and to avoid generating ambivalent code, we demand that all syntactical language elements of the same type have unique names.

## U1: Automata within a component have unique names.

Automata with duplicate names inside a common parent component, as seen in Listing 6.1, are not permitted.

```
                                                              MontiArcAutomaton
 1  component IntegerBufferU1 {
 2
 3    automaton BufferAutomaton {
 4      // ...
 5    }
 6
 7    automaton BufferAutomaton { ⊗ // duplicate automaton definition
 8      // ...
 9    }
10  }
```

Listing 6.1: Violated CoCo U1: Two automata definitions with equal name

## U2: State names are unique within an automaton.

All state names in an automaton have to be unique in the scope of the automaton definition. This requirement holds for state names listed in a single `state` definition as well as among multiple `state` statements.

```
                                                              MontiArcAutomaton
 1  component IntegerBufferU2 {
 2
 3    automaton BufferAutomaton {
 4      state S, T, S; ⊗ // state 'S' defined multiple times
 5      state T;       ⊗ // state 'T' defined earlier
 6    }
 7  }
```

Listing 6.2: Violated CoCo U2: Duplicate state definitions

In Listing 6.2 the condition U2 is violated by declaring a state named `S` twice in a single `state` definition (l. 4). Furthermore, both state definitions declare a state with the name `T` (ll. 4-5), which also violates this condition.

## U3: The names of variables and ports are unique within a component.

As all variables are declared and referenced in the scope of a component, the name of each variable must be different from all other variable names and port names. This holds for variables of the same type as well as for variables of different types.

```
                                                    MontiArcAutomaton
 1  component IntegerBufferU3 {
 2
 3    port
 4      in Integer input;
 5
 6    String input;  ⊗ // port with name 'input' already exists
 7
 8    Integer buffer;
 9    String buffer;  ⊗ // variable 'buffer' defined twice
10  }
```

Listing 6.3: Violated CoCo U3: Duplicate variable definitions

In Listing 6.3 context condition U3 is violated twice: first, by defining a variable with a name already assigned to a port (l. 6) and second, by duplicate declaration a of variables with name `buffer`.

## 6.2. Convention Conditions

We introduce a set of conventions for MontiArcAutomaton models. A model that violates a convention rule can still be considered well-formed but is strongly discouraged. Violations of the rules introduced in this section result in warnings instead of errors.

### C1: An automaton has at least one initial state.

At least one state of each automaton has to be declared as an initial state. In Listing 6.4 this context condition is violated.

```
                                                    MontiArcAutomaton
 1  component IntegerBufferC1 {
 2
 3    automaton BufferAutomaton { ⚠ // initial state missing
 4      state S;
 5    }
 6  }
```

Listing 6.4: Violated CoCo C1: Missing initial state declaration

### C2: The names of variables and ports start with lowercase letters.

In order to conform to conventions known from Java we discourage the use of port and variable names which begin with an uppercase letter (see Listing 6.5).

```
                                                    MontiArcAutomaton
1  component IntegerBufferC2to4 {
2    port
3      in Integer Input;  ⚠ // port name must start lowercase
4
5    automaton buffer {  ⚠ // automaton name must start uppercase
6      state s;  ⚠ // state name must start uppercase
7    }
8  }
```

Listing 6.5: Violated CoCos C2, C3, C4: Variable, automata, and state names are not defined in compliance to MontiArcAutomaton conventions

### C3: The names of automata start with uppercase letters.

An automaton is a constant entity and should thus have a name which begins with an uppercase letter (see Listing 6.5).

### C4: The names of states start with uppercase letters.

An automaton's states are static entities and have therefore names that start in uppercase (see Listing 6.5).

## 6.3. Referential Integrity Conditions

This section introduces rules for the well-definedness of references to language elements inside MontiArcAutomaton models.

### R1: States referenced by a transition must be declared.

The states which represent the source or the target of a transition have to be declared in a `state` statement. Two violations of this context condition are depicted in Listing 6.6. The state `T` has not been declared but is referenced as a target state in line 9 and as a source state in line 10.

### R2: Ports and variables referenced on transitions must be declared.

Only ports that have been declared as input or output ports in the components interface may be referenced by a transition to either send or receive messages. In Listing 6.7 the ports or variables with names `saveValue`, `buffer`, `input`, and `output` are unknown to both component and automaton. As they are used to receive and send messages (l. 13), the automaton definition is erroneous with respect to R2. The same holds for reading and assigning variables. Additionally, ports and variables that are used as messages or values must be declared as well.

```MontiArcAutomaton
1  component IntegerBufferR1 {
2
3    port
4      in Boolean saveValue;
5
6    automaton BufferAutomaton {
7      state S;
8
9      S -> T true;  ⊗ // state 'T' is undefined
10     T -> S false; ⊗ // state 'T' is undefined
11   }
12 }
```

Listing 6.6: Violated CoCo R1: Reference to an undefined state

```MontiArcAutomaton
1  component IntegerBufferR2 {
2
3    automaton BufferAutomaton {
4      state S;
5
6      S saveValue = true / {buffer = input, output = 0 };
7      ⊗ // name 'saveValue' is undefined
8      ⊗ // name 'buffer' is undefined
9      ⊗ // name 'input' is undefined
10     ⊗ // name 'output' is undefined
11   }
12 }
```

Listing 6.7: Violated CoCo R2: Multiple undefined ports and variables

### R3: Variable declarations may not reference ports.

As a convention, we require that initial value assignments to variables are performed prior to any communication taking place. Thus, variable declarations may not reference any ports.

## 6.4. Type Correctness Conditions

This section summarizes rules for the correct usage and combination of typed elements in MontiArcAutomaton models.

**T1: Messages sent or received via ports and read from or assigned to variables must conform to the according port or variable types.**

For every port and variable in every input and output block, messages must consist solely of values that conform to the type of the port and variable respectively.

```
                                                        MontiArcAutomaton
 1  component IntegerBufferT1 {
 2
 3    port
 4      in Integer input,
 5      in Boolean saveValue,
 6      out Integer output;
 7
 8    Integer buffer;
 9
10    automaton BufferAutomaton {
11      state S;
12
13      S -> S true /  {buffer = true, output = "Zero" };
14      ❌ // 'true' is no Integer
15      ❌ // 'Zero' is no Integer
16    }
17  }
```

Listing 6.8: Violated CoCo T1: Using incorrectly typed values with ports or variables

Listing 6.8 demonstrates two violations of the context condition T1: The port `buffer` of type `Integer` cannot be used to send a message of the type `Boolean` (l. 13), and a value of the type `String` cannot be assigned to the port `output` (l. 13). This context condition applies to input and output blocks of transitions as well as to output blocks of initial state outputs.

**T2: Initial values of variables must conform to their types.**

The possibility of assigning an initial value to a variable is provided by the underlying grammar. For initial assignments the same rules apply as demanded by the context condition T1. An assigned value must conform to the type of the variable. A violation of the context condition T2 can be seen in Listing 6.9. In line 3 a literal of the type `String` cannot be assigned to a variable of the type `Integer`.

**T3: Input ports and variables used as part of a message or assignment, must conform to the according port and variable types.**

When using references to input ports or variables as messages on transitions, their types must be compatible with the type of the port or variable on the left-hand side of the comparison or assignment. Two possible mistakes are demonstrated in Listing 6.10, l. 13.

```
   ┌────────────────────────────────────────── MontiArcAutomaton ──┐
 1 │component IntegerBufferT2 {
 2 │
 3 │   Integer buffer = "Hello";  ⊗ // 'Hello' is no Integer
 4 │
 5 │   automaton BufferAutomaton {
 6 │      state S;
 7 │      //...
 8 │   }
 9 │}
   └────────────────────────────────────────────────────────────────┘
```

Listing 6.9: Violated CoCo T2: Assigning incompatible values in variable declaration
statements

```
   ┌────────────────────────────────────────── MontiArcAutomaton ──┐
 1 │component EchoT3 {
 2 │   port
 3 │      in String input,
 4 │      in Boolean speak,
 5 │      out String output;
 6 │
 7 │   Boolean tmp;
 8 │
 9 │   automaton {
10 │      state S;
11 │      initial S;
12 │
13 │      S true / tmp = input, output = ["input is:", speak];
14 │      ⊗ // port 'input' is no Boolean
15 │      ⊗ // variable 'speak' is no String
16 │   }
17 │}
   └────────────────────────────────────────────────────────────────┘
```

Listing 6.10: Violated CoCo T3: Constructing a message sequence with an invalid
concatenation of messages of incompatible types and assigning a message
of an incompatible type to a variable (l. 13)

## T4: The special literal value `NoData` cannot be used with variables.

The value `NoData` (written `--`) is reserved for timed streams on ports as it represents
the absence of a message in a time slice. Listing 6.11 demonstrates all kinds of variable
accesses that are prohibited if `NoData` is used as a value.

```
                                                    ┌──────────────────────┐
                                                    │ MontiArcAutomaton    │
 1 component IntegerBufferT4 {
 2
 3   port
 4     in Integer input,
 5     out Integer output;
 6
 7   Integer buffer = --;
 8   ❌ // cannot assign NoData to variable 'buffer'
 9
10   automaton BufferAutomaton {
11     state S;
12
13     S -> S buffer = -- / buffer = --;
14     ❌ // cannot read NoData from variable 'buffer'
15     ❌ // cannot write NoData to variable 'buffer'
16   }
17 }
```

Listing 6.11: Violated CoCo T4: Assigning and reading `NoData` from variables

## T5: Sequences of values cannot be read from or assigned to variables.

Variables can hold only single values. Therefore, they can neither be assigned sequences of values nor can they be queried for such sequences (see Listing 6.12).

```
                                                    ┌──────────────────────┐
                                                    │ MontiArcAutomaton    │
 1 component IntegerBufferT5 {
 2
 3   Integer buffer;
 4
 5   automaton {
 6     state S;
 7     initial S;
 8
 9     S buffer = [1, 0] / buffer = [1, 1];
10     ❌ // cannot read sequence from variable
11     ❌ // cannot write sequence to variable
12   }
13 }
```

Listing 6.12: Violated CoCo T5: Writing and reading sequences of values from variables

**T6: The direction of ports has to be respected.**

Messages can only be received through input ports and sent only through output ports. Hence, transitions have to obey these limitations by only reading from input ports and reacting by sending messages via output ports. Due to context condition T6 transitions may not use ports in the opposite direction as illustrated in Listing 6.13.

```
                                                          MontiArcAutomaton
1  component IntegerBufferT6 {
2
3    port
4      in Integer input,
5      out Integer output;
6
7    automaton BufferAutomaton {
8      state S;
9
10     S output = 0 / input = 1;
11  ❌ // receiving from output port 'output'
12  ❌ // sending to input port 'input'
13   }
14 }
```

Listing 6.13: Violated CoCo T6: Violation of port directions

**T7: Output ports must not be used as part of messages in a transition's input or output block.**

References to input ports can be used for forwarding received messages on output ports. In this case the currently received value at the input port is forwarded as (a part of) the output. For output ports there are no current values which could be accessed and thus it is not allowed to reference output ports when constructing messages. Listing 6.14 demonstrates two different violations of context condition T7.

## 6.5. MAA$_{ts}$ Specific Context Conditions

Some context conditions for MontiArcAutomaton models are specific to language profiles, e.g., the language profile for time-synchronous communication with strong causality (MAA$_{ts}$ described in Section 4.1). We list these context conditions here.

**S1$_{ts}$: An atomic component contains at most one automaton.**

The language profile MAA$_{ts}$ does not allow the definition of two or more automata within the same component. Therefore plural automata declarations, as seen in Listing 6.15, are

```
                                                              MontiArcAutomaton
1  component IntegerBufferT7 {
2
3    port
4      in Boolean saveValue,
5      out Integer output;
6
7    Integer buffer;
8
9    automaton BufferAutomaton {
10     state S;
11
12     S true / buffer = output;
13     ❌// output port 'output' used in message
14     S buffer = output;
15     ❌// output port 'output' used as value
16   }
17 }
```

Listing 6.14: Violated CoCo T7: Use of an output port in messages of a transition's
                output block

not permitted. Other language profiles can be defined to handle multiple and concurrent
automata.

```
                                                              MontiArcAutomaton-TS
1  component IntegerBufferS1 {
2
3    automaton FirstBuffer {
4      // ...
5    }
6
7    automaton SecondBuffer { ❌// multiple automata not
8                               // allowed in this profile
9      // ...
10   }
11 }
```

Listing 6.15: Violated CoCo S1$_{ts}$ for profile MAA$_{ts}$: Multiple automata definitions

## S2$_{ts}$: Ports must not be used as part of messages in initial state outputs.

Strong causality introduces a computation delay, i.e., an output may not depend on the
input received at the same point in time. Thus, an initial output may not depend on
values read from input ports. In Listing 6.16 it is shown that it is strictly forbidden to reference ports within an initial state output. This context condition applies to the language profile MAA$_{ts}$.

```
                                                    MontiArcAutomaton-TS
 1  component IntegerBufferS2TS {
 2
 3    port
 4      in Integer input,
 5      out Integer output;
 6
 7    automaton BufferAutomaton {
 8      state S, T;
 9
10      initial S / output = input;
11      ❌ // port in message for initial state output
12    }
13  }
```

Listing 6.16: Violated CoCo S2$_{ts}$ for profile MAA$_{ts}$: A port as message source in an initial state output

## S3$_{ts}$: In every cycle at most one message per port is sent.

The language profile MAA$_{ts}$ does not allow emitting more than one message per port in a cycles. Listing 6.17 demonstrates a violation by sending a sequence of messages. This context condition applies to the language profile MAA$_{ts}$.

```
                                                    MontiArcAutomaton-TS
 1  component InitialFib {
 2
 3    port
 4      in Integer input,
 5      out Integer output;
 6
 7    automaton {
 8      state S;
 9      initial S / output = [0, 1, 1, 2, 3, 5, 8, 13];
10      ❌ // sending sequence of messages not allowed
11    }
12  }
```

Listing 6.17: Violated CoCo S3$_{ts}$ for profile MAA$_{ts}$: Multiple messages per port

## 6.6. MAA$_{ed}$ Specific Context Conditions

Some context conditions for MontiArcAutomaton models are specific to language profiles, e.g., the language profile for event-driven communication (MAA$_{ed}$ described in Section 4.2). We list these context conditions here.

### S1$_{ed}$ for profile MAA$_{ed}$: An atomic component contains at most one automaton.

The language profiles MAA$_{ed}$ does not allow the definition of two or more automata within the same component. Therefore plural automata declarations, as seen in Listing 6.18, are not permitted. Other language profiles can be defined to handle multiple and concurrent automata.

```
                                                      MontiArcAutomaton-ED
 1  component IntegerBufferS1 {
 2
 3    automaton FirstBuffer {
 4      // ...
 5    }
 6
 7    automaton SecondBuffer { ❌ // multiple automata not
 8                                // allowed in this profile
 9      // ...
10    }
11  }
```

Listing 6.18: Violated CoCo S1$_{ed}$: Multiple automata definitions

### S2$_{ed}$: All inputs must be processed one single message at a time.

The language profile for event-driven automata (see Section 4.4) demands that a transition is triggered by a single message on a single port. This is equivalent to various applications within embedded systems where events are represented by interrupts and routines handling them. For that reason it is not allowed for a transition to consume multiple messages at once. Listing 6.19 demonstrates a violation of context condition S2$_{ed}$ in two different ways. In line 11 two input ports are read in a guard expression. Likewise, line 13 violates the condition by simultaneously reading two messages from different ports directly.

```
    ┌─────────────────────────────────────────────── MontiArcAutomaton-ED ┐
 1  │ component Filter {
 2  │
 3  │   port
 4  │     in Integer input,
 5  │     in Integer threshold,
 6  │     out Integer output;
 7  │
 8  │   automaton BufferAutomaton {
 9  │     state S;
10  │
11  │     S [ocl: input > threshold] / output = input;
12  │       ⊗ // reading from multiple inputs inside a guard
13  │     S input = 1, threshold = 1 / output = -1;
14  │       ⊗ // reading from multiple ports
15  │   }
16  │ }
    └──────────────────────────────────────────────────────────────────────┘
```

Listing 6.19: Violated CoCo S2$_{ed}$ for profile MAA$_{ed}$: Reading multiple messages

## S3$_{ed}$: The −− symbol may not be used as input in MAA$_{ed}$ automata.

The language profile for event-driven automata (see Section 4.4) does not allow triggering reactions by the absence of events. A violation is illustrated in Listing 6.20, l. 11.

```
    ┌─────────────────────────────────────────────── MontiArcAutomaton-ED ┐
 1  │ component TemperatureMonitor {
 2  │   port
 3  │     in Integer temp,
 4  │     out String message;
 5  │
 6  │   automaton {
 7  │     state   S;
 8  │     initial S;
 9  │
10  │     S  x = 0  / cout = "freezing";
11  │     S  x = -- / cout = "nothing happened";
12  │       ⊗   // triggered by absence of an event
13  │   }
14  │ }
    └──────────────────────────────────────────────────────────────────────┘
```

Listing 6.20: Violated CoCo S3$_{ed}$ for profile MAA$_{ed}$: Using −− as trigger

# Chapter 7.

# Case Studies

We have evaluated MontiArcAutomaton on several platforms from simulators to educational Lego NXT robots to complex service robots in a distributed robotic system. To evaluate MontiArcAutomaton on these platforms, we have developed code generators to different target languages [RRW13b], which include code generation to Mona [EKM98] (formal analysis), EMF Ecore[1] (graphical editing), and Java and Python (deployment). The code generators are generic in the sense, that they generate implementations of composed components and automata in their target language. Many applications require additional components to access platform specific software and hardware. MontiArcAutomaton libraries organize these models and their platform specific implementations

- for robots using ROS [QCG+09],

- for robots using SmartSoft [SSL11],

- for Lego NXT robots using leJOS[2], and

- for simulators such as ROS turtlesim[3] and Simbad[4].

The libraries comprise from 4 (ROS turtlesim) to 21 (leJOS NXT) components and can be easily imported by MontiArcAutomaton applications to use these with different platforms. Experiences with leJOS and ROS have shown that developing a library for a certain platform is relatively straightforward as the functionality wrapped per component is usually well defined by existing APIs. Additionally, we have developed GPL specific libraries to provide GPL functionalities (e.g., file I/O). We have evaluated MontiArc-Automaton in different projects and lab courses[5] on various platforms.

Figure 7.1 depicts a Lego NXT robot, which is controlled by the bumper bot software architecture shown in Figure 3.2. All components of the software architecture are deployed to the NXT computation unit, which is connected to an ultrasonic sensor and two motors — one for each track. Please note how close the logical architecture reflects the physical implementation in this case.

---

[1]The Eclipse Modeling Framework Project: http://www.eclipse.org/modeling/emf/
[2]leJOS website: http://lejos.sourceforge.net/
[3]ROS turtlesim website: http://wiki.ros.org/turtlesim
[4]Simbad website: http://simbad.sourceforge.net/
[5]Videos of results of these courses are available at http://monticore.de/robotics/.

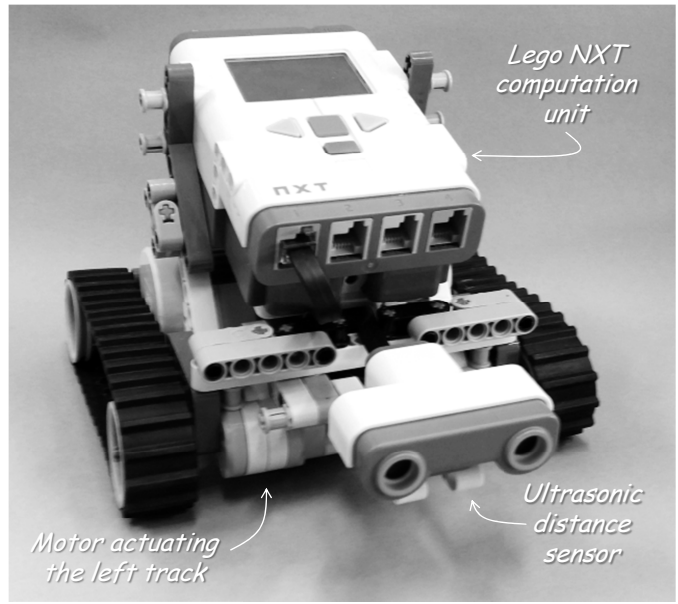

Figure 7.1.: A Lego NXT robot controlled by the bumper bot software architecture of Figure 3.2

## 7.1. Lego NXT Coffee Service

With MontiArcAutomaton, the Java code generator, and the Java leJOS NXT libraries, we evaluated the MontiArcAutomaton framework during a university lab course in the winter term 2012/13 with a eight master level students [RRW13a]. The students modeled the software architecture for a distributed robotic coffee service consisting of the three robots illustrated in Figure 7.2. The GPL Java was used to implement the behavior of components not easily expressible as automata (two out of ten component models).

The system enables users to issue requests for coffee via a website hosted on a smart phone. This phone is connected to the coffee preparation robot via Bluetooth. Once this robot receives a request, it informs the coffee delivery robot to fetch a plastic mug from the mug provider robot. Afterwards, it returns to the coffee preparation robot, instructs it via Bluetooth to make coffee and afterwards drives to the user who ordered the coffee using — in the absence of sophisticated localization sensors — black lanes with colored junctions.

## 7.2. ROS Robotino Logistics

In winter term 2013/14, we evaluated MontiArcAutomaton with a Python code generator and the ROS Python Robotino modules in another lab course. In this course, nine master level students developed a model-driven logistics application using a Robotino robot with a Kinect for user interaction. The software architecture was modeled with 31 components. Of these, nine contain automata and 17 have platform specific implementations. The top-level of this architecture is depicted in Figure 7.3. The components `Navigation`, `MapProvider`, and `TaskManager` are composed from other components. During this course, the students modeled the software architecture and programmed component GPL implementations where necessary (as, for instance, the behavior of the ROS-specific components wrapping the Kinect had to be implemented manually).

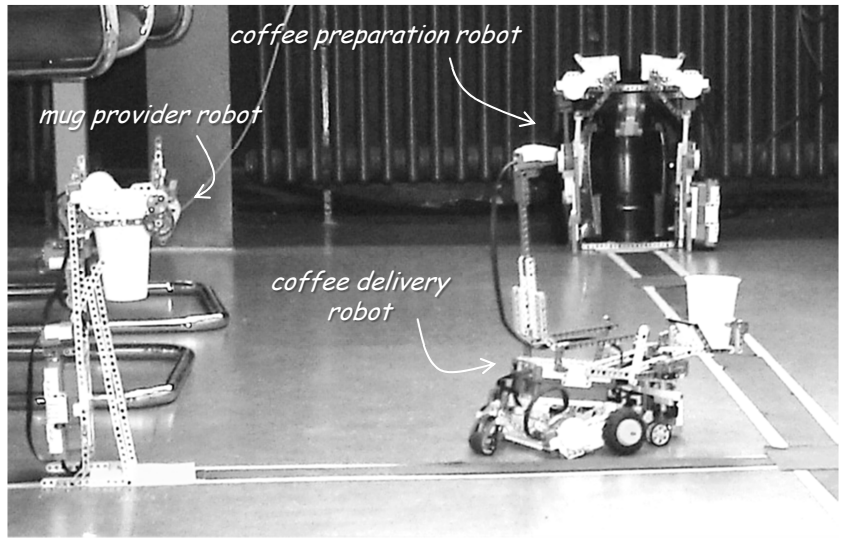

Figure 7.2.: The distributed robotic coffee service as implemented with Lego NXT robots
running leJOS as target platform

Here, a web server is hosted on the robot itself and receives tasks such as "fetch item X from room Y". Automata translate these tasks into motion and interaction commands passed to components `Navigation` and `UserInterface` which translate these into platform-specific primitives.

## 7.3. SmartSoft Robotino Logistics

In a lab course of summer term 2014 we assigned the task to develop a robotics logistics application similar to the previous one. 14 students from different computer science bachelor and master programs participated. The students modeled the software architecture with MontiArcAutomaton and connected it to the SmartSoft [SSL11] middleware to control the robot. For his, they produced the architecture depicted in Figure 7.4. Of the depicted subcomponents, five are composed and two atomic.

To communicate with the robot, both a website and a tablet PC were used. Both are connected to the architecture via subcomponents of component `Backend`. Logistics tasks are passed from `Backend` to `JobManager` and translated into commands send to the SmartSoft middleware via component `SequencerProxy`.

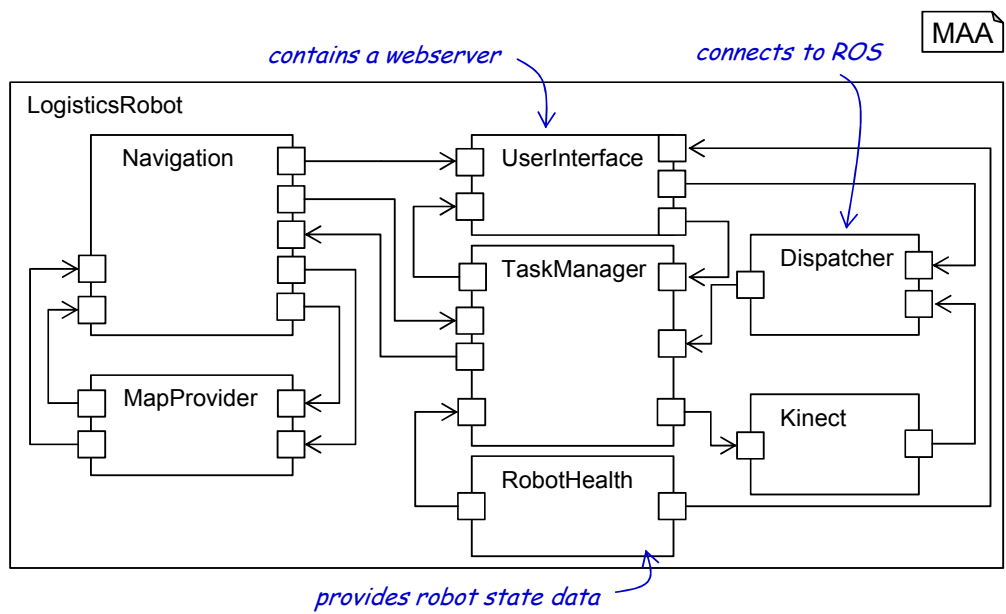

Figure 7.3.: The top-level architecture of the logistics application implemented during
        winter term 2013/14 for a Robotino running ROS and Python

## 7.4. IServeU

IServeU is a research project of various academic and professional partners revolving
around the engineering and deployment of model-driven robotics applications to real-
world contexts. In this 3-year project, funded by the German Federal Ministry of Ed-
ucation and Research (BMBF), MontiArcAutomaton is used to model the architecture
and parts of the behavior for a logistics robotics application in a complex environment.
MontiArcAutomaton serves as ADL for a high-level controller which again interfaces
SmartSoft.

Figure 7.5 shows the core component `RobotController` of the IServeU top-level
software architecture. The `RobotController` receives tasks and passes these to the
`Scheduler` which decomposes them into individual goals passed to a task planner capa-
ble of reasoning about reaching goals based on the current situation and actions available
to the robot. To deduct a valid plan it may read properties of the robot and its environ-
ment via component `PropertyCalculator`. Once a plan is deducted, the controller
executes its actions via component `ActionExecuter`, which is connected to the under-
lying SmartSoft middleware.

All components types depicted are composed from multiple subcomponents and the
software architecture comprises 20 different component types. Of these, seven are com-
posed from other components and eight are atomic components containing automata.
The behavior of two components is generated from high-level models of the robot and
its capabilities.

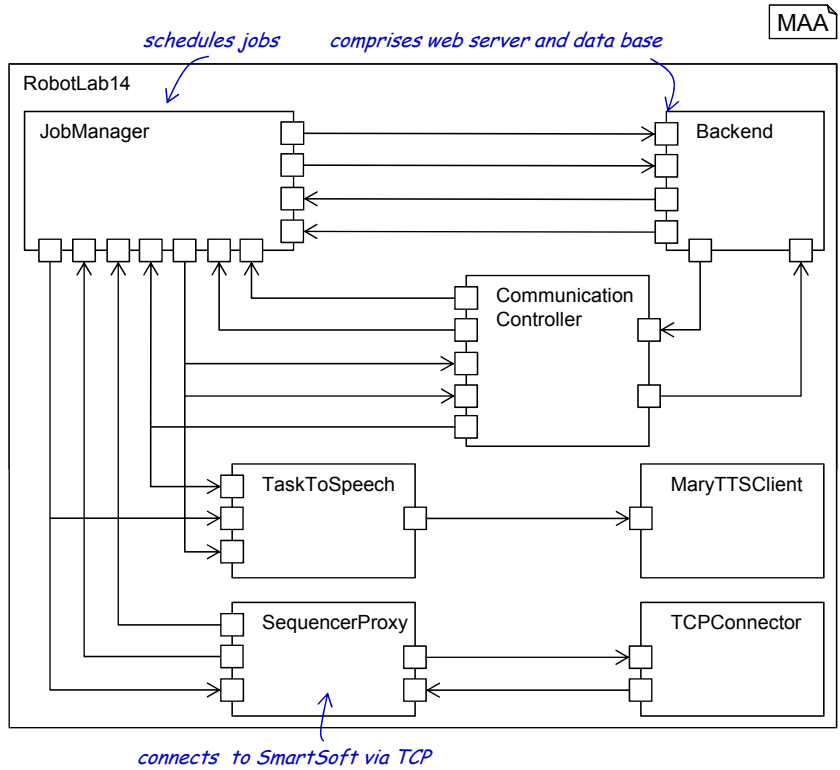

Figure 7.4.: Top-level architecture of the logistics application implemented in summer term 2014 for a Robotino running SmartSoft and Python. Five of the displayed components types are composed

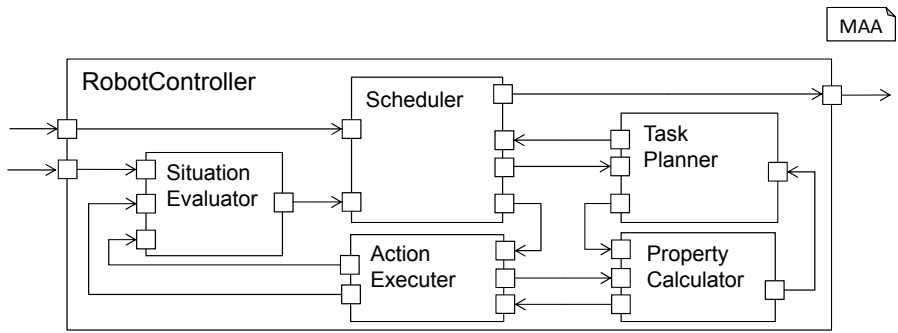

Figure 7.5.: The top-level architecture of the high-level robot MontiArcAutomaton controller employed in the IServeU project

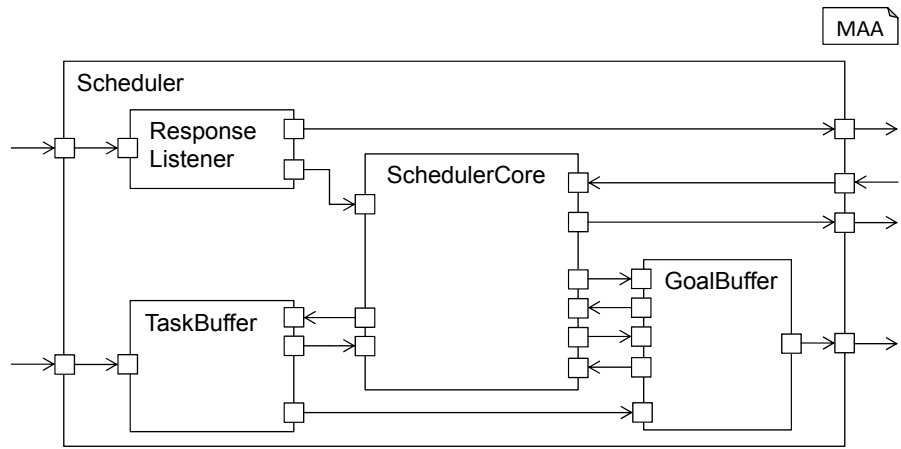

Figure 7.6.: The software architecture of component type `Scheduler` responsible for managing tasks and goal execution. The subcomponent `SchedulerCore` contains an automaton

The decomposed component `Scheduler`, depicted in Figure 7.6 takes care of overall task scheduling. For this, the automaton in subcomponent `SchedulerCore` breaks down tasks into individual goals and takes care of the remaining goals using the connected buffer components.

# Chapter 8.

# Discussion and Current Limitations

Based on the lessons learned from past and ongoing evaluation of the MontiArcAutomaton language and framework we discuss its possible extensions. The three main topics we address concern future directions of research around MontiArcAutomaton regarding the expressiveness of the language, its extensibility with additional behavior modeling languages, and the code generation and deployment to different robotics platforms.

## 8.1. Expressiveness of Automata for Behavior Modeling

We have evaluated the modeling language MontiArcAutomaton with groups of students regarding its expressiveness for modeling the behavior of robotic systems as described in Chapter 7. Regarding expressiveness of the behavior modeling part of MontiArcAutomaton we found out that some components are not easily modeled in MontiArcAutomaton. For example, the language is missing suitable means for modeling (dynamic) look-up tables and more complex algorithms, e.g., for computing shortest paths in a graph. Another limitation is the lack of expressiveness for message data manipulation in MontiArcAutomaton. Examples are the computation of arithmetic or logical expressions when assigning values to variables and sending messages on ports. We are currently working on extending these assignments to allow expressions similar to the OCL and Java expressions evaluated in guards.

To overcome current limitation of the language expressiveness in [RRW13a, RRW13b] we allowed native component implementations in programming languages such as Java and Python.

Expressing further properties of component behavior common in CPS requires more fundamental extensions of the automaton language. Many formalisms for modeling automata with (multiple) clocks, differential equations, and probabilities [Rab63, AD94, GSB98, LSV03, LS11] exist in the literature and could be integrated into MontiArcAutomaton. The semantics of MontiArcAutomaton is based on the FOCUS framework with powerful refinement and composition operators. Extensions the system model of FOCUS that support hybrid and dense-timed streams have been suggested in [SRS99]. We thus believe that the current MontiArcAutomaton can seamlessly be extended to model advanced features of CPS.

## 8.2. Extension with Additional Behavior Modeling Languages

One of the main concepts of MontiArcAutomaton is the composition of components from subcomponents with well-defined interfaces. The encapsulation of functionality in components and the hiding of implementation details not relevant for composition allow logically distributed development and physically distributed computation models. Encapsulation of component behavior and its composition in a uniform way following the FOCUS calculus makes MontiArcAutomaton amenable for extending component models with additional behavior description languages complementary to automata.

Examples for other component behavior modeling languages are simple, stateless I/O rules [RRW13c] or arbitrary domain specific languages defined using the MontiCore framework. We have presented extensions of MontiArcAutomaton with additional component behavior modeling languages and underlying concepts in [RRW13c, LPR+13].

## 8.3. Code Generation, Deployment, and Platforms

MontiArcAutomaton is a modeling language for architecture and behavior modeling of CPS. One application domain of MontiArcAutomaton are robotic systems. The transition from platform independent models to concrete platform dependent applications executable on systems of physical robots is a challenge to generative software engineering.

We have developed code generators for target languages including EMF, Java, Mona, and Python and to target platforms executing leJOS, ROS, and SmartSoft environments (see Chapter 7 and [RRW13b, RRW13c]). Current challenges in the efficient development of code generators are developing and applying concepts for compositional code generation as well as reusing (parts of) generators, models, and manual implementations for different target platforms. We report on an approach for reusing models and manual implementations organized in libraries in [RRW14].

# Chapter 9.

# Summary

In this work we presented a modeling language for the description of software architectures of Cyber Physical Systems as Component and Connector software architecture models. The modeling language MontiArcAutomaton extends the Architecture Description Language MontiArc and inherits all language features described in [HRR12]. MontiArcAutomaton thus allows the modeling of components with well-defined interfaces consisting of typed ports. Components are either modeled as the hierarchical composition of subcomponents interacting by exchanging messages via directed connectors between ports or as atomic components. The language integrates syntactical elements of I/O$^\omega$ automata into atomic component definitions to model the interaction behavior of components.

MontiArcAutomaton is a modeling language which allows the definition of language profiles to express automata for various models of computation of interactive systems. We sketched two language profiles and their semantics in Chapter 4. The syntax of MontiArcAutomaton automata is introduced in a comprehensive language reference in Chapter 5. The language reference covers all syntactical features added to the MontiArc language. These include automata with states and transitions that depend on the values of local variables and messages received on the input ports of a component. Components interact by sending messages via their output ports. MontiArcAutomaton also comprises well-formedness rules for models to ensure valid references and types. Their complete list was presented in Chapter 6. Examples for these rules are the existence of ports and variables referenced on transitions or the type compatibility of assigned values and messages sent.

One advantage of MontiArcAutomaton is its ability for modeling requirements for component behavior as well as component implementations [RRW12]. An application domain are robotics applications [RRW13c]. For this, we have implemented code generators for MontiArcAutomaton models to various target languages including executable code for robotics platforms [RRW13b]. The modeling language MontiArcAutomaton and its code generation framework were evaluated multiple case studies (Chapter 6) which led to identification of issues and possibilities for future extension (Chapter 8).

# List of Figures

# List of Listings

# Appendix A.

# Human Readable Grammar

The MontiArcAutomaton grammar shown in Listing A.1 is provided for human comprehension in EBNF-like style. It is a simplified version of the MontiCore grammar used to create MontiArcAutomaton models and describes the concrete syntax only.

MontiArcAutomaton extends the MontiArc grammar (l. 3), which is given in the appendix of [HRR12]. The MontiArcAutomaton grammar introduces productions `Automaton` (ll. 5-7) and `VariableDeclaration` (l. 9) and the productions used by these. Both, `Automaton` and `VariableDeclaration`, can be used inside components independently.

The `Automaton` production rule is composed from production rules for `States` (l. 13), `InitialStates` (l. 17), and `Transition` (l. 19). These contain the productions for the language elements described in Chapter 5.

```
                                                    ┌─────────────────────────┐
                                                    │    MontiCore Grammar     │
 1  package mc.maautomaton;
 2
 3  grammar MontiArcAutomaton extends MontiArc {
 4
 5    Automaton =
 6      Stereotype? "automaton" Name? "{"
 7        (States | InitialStates | Transition)* "}";
 8
 9    Variables = "var"? Type (VariableAssignment||",")+ ";";
10
11    VariableAssignment = Name ("=" Value)?;
12
13    States = "state" (State||",")+ ";" ;
14
15    State = Stereotype? Name;
16
17    InitialStates = "initial" (Name||",")+ ("/" Output)? ";";
18
19    Transition =
20      source:Name ("->" target:Name)?
21      Guard? Input? ("/" Output)? ";";
22
23    Guard = "[" InvariantContent "]";
```

```
24
25   Input = "{" MatchList "}" | MatchList;
26
27   MatchList = (Match||",")+;
28
29   Match = (Name "=")? (OptionalValue||"|")+;
30
31   OptionalValue = Value | "--";
32
33   Output = "{" AssignmentList "}" | AssignmentList;
34
35   AssignmentList = (Assignment||",")+;
36
37   Assignment = (Name "=")? (OptionalValueOrSequence||"|")+;
38
39   OptionalValueOrSequence = OptionalValue | ValueSequence;
40
41   ValueSequence = "[" (Value||",")* "]";
42 }
```

Listing A.1: The complete MontiArcAutomaton grammar for human readers

# Appendix B.

# Parser Grammar

The MontiArcAutomaton grammar in Listing B.1 is used as input for the MontiCore tool to parse MontiArcAutomaton models and create Abstract Syntax Trees (AST). It contains additional information for the MontiCore parser and lexer framework (ll. 5-9) as well as specific rules to improve parsing. Integration of new language elements into MontiArc components in enabled via the interface `ArcElement` (cf. [HRR12]), which both `Automaton` and `VariableDeclaration` implement (ll. 11 and 18). The production for the `--` terminal similarly implements the interface `Value` which enables to use `--` within guard expressions as well.

```
                                                    ┌─── MontiCore Grammar ───
1  package mc.maautomaton;
2
3  grammar MontiArcAutomaton extends mc.umlp.arc.MontiArc {
4
5    options {
6      compilationunit ArcComponent
7      parser lookahead=5
8      lexer lookahead=7
9    }
10
11   Automaton implements (Stereotype? "automaton" Name?) =>
12     ArcElement =
13       Stereotype? "automaton" Name? "{"
14         (States | InitialStates | Transition)* "}";
15
16   NoData implements ("--") => Value = "--";
17
18   Variables = "var"? Type
19     VariableAssignment ("," VariableAssignment)* ";";
20
21   VariableAssignment = Name ("=" Value)?;
22
23   States = "state" State ( "," State )* ";" ;
24
25   State = Stereotype? Name;
26
```

```
27   InitialStates = "initial" Name ("," Name)* ("/" Output)? ";";
28
29   Transition =
30     source:Name ("->" target:Name)?
31     Guard? Input? ("/" Output)? ";";
32
33   Guard = "[" (kind:Name ":")? InvariantContent "]";
34
35   Input = "{" MatchList "}" | MatchList;
36
37   MatchList = Match ("," Match)*;
38
39   Match = (Name "=")? OptionalValue ("|" OptionalValue)*;
40
41   OptionalValue = Value | NoData;
42
43   Output = "{" AssignmentList "}" | AssignmentList;
44
45   AssignmentList = Assignment ("," Assignment)*;
46
47   Assignment = (Name "=")?
48     (OptionalValueOrSequence ("|" OptionalValueOrSequence))*;
49
50   OptionalValueOrSequence = OptionalValue | ValueSequence;
51
52   ValueSequence = "[" (Value||",")* "]";
53
54   ValueSequence = "[" (Value ("," Value)*)? "]";
55
56 }
```

Listing B.1: The MontiArcAutomaton grammar for parsing

# Related Interesting Work from the SE Group, RWTH Aachen

### Agile Model Based Software Engineering

Agility and modeling in the same project? This question was raised in [Rum04]: "Using an executable, yet abstract and multi-view modeling language for modeling, designing and programming still allows to use an agile development process." Modeling will be used in development projects much more, if the benefits become evident early, e.g with executable UML [Rum02] and tests [Rum03]. In [GKRS06], for example, we concentrate on the integration of models and ordinary programming code. In [Rum12] and [Rum11], the UML/P, a variant of the UML especially designed for programming, refactoring and evolution, is defined. The language workbench MontiCore [GKR+06] is used to realize the UML/P [Sch12]. Links to further research, e.g., include a general discussion of how to manage and evolve models [LRSS10], a precise definition for model composition as well as model languages [HKR+09] and refactoring in various modeling and programming languages [PR03]. In [FHR08] we describe a set of general requirements for model quality. Finally [KRV06] discusses the additional roles and activities necessary in a DSL-based software development project.

### Generative Software Engineering

The UML/P language family [Rum12, Rum11] is a simplified and semantically sound derivate of the UML designed for product and test code generation. [Sch12] describes a flexible generator for the UML/P based on the MontiCore language workbench [KRV10, GKR+06]. In [KRV06], we discuss additional roles necessary in a model-based software development project. In [GKRS06] we discuss mechanisms to keep generated and handwritten code separated. In [Wei12] we show how this looks like and how to systematically derive a transformation language in concrete syntax. To understand the implications of executability for UML, we discuss needs and advantages of executable modeling with UML in agile projects in [Rum04], how to apply UML for testing in [Rum03] and the advantages and perils of using modeling languages for programming in [Rum02].

### Unified Modeling Language (UML)

Many of our contributions build on UML/P described in the two books [Rum11] and [Rum12] are implemented in [Sch12]. Semantic variation points of the UML are discussed in [GR11]. We discuss formal semantics for UML [BHP+98] and describe UML semantics using the "System Model" [BCGR09a], [BCGR09b], [BCR07b] and [BCR07a]. Semantic variation points have, e.g., been applied to define class diagram semantics [CGR08]. A precisely defined semantics for variations is applied, when checking variants of class diagrams [MRR11c] and objects diagrams [MRR11d] or the consistency of both kinds of diagrams [MRR11e]. We also apply these concepts to activity diagrams (ADs) [MRR11b] which allows us to check for semantic differences of activity diagrams [MRR11a]. We also discuss how to ensure and identify model quality [FHR08], how models, views and the system under development correlate to each other [BGH+98] and how to use modeling in agile development projects [Rum04], [Rum02] The question how to adapt and extend the UML in discussed in [PFR02] on product line annotations for UML and to more general discussions and insights on how to use meta-modeling for defining and adapting the UML [EFLR99], [SRVK10].

## Domain Specific Languages (DSLs)

Computer science is about languages. Domain Specific Languages (DSLs) are better to use, but need appropriate tooling. The MontiCore language workbench [GKR+06], [KRV10], [Kra10] describes an integrated abstract and concrete syntax format [KRV07b] for easy development. New languages and tools can be defined in modular forms [KRV08, Völ11] and can, thus, easily be reused. [Wei12] presents a tool that allows to create transformation rules tailored to an underlying DSL. Variability in DSL definitions has been examined in [GR11]. A successful application has been carried out in the Air Traffic Management domain [ZPK+11]. Based on the concepts described above, meta modeling, model analyses and model evolution have been examined in [LRSS10] and [SRVK10]. DSL quality [FHR08], instructions for defining views [GHK+07], guidelines to define DSLs [KKP+09] and Eclipse-based tooling for DSLs [KRV07a] complete the collection.

## Modeling Software Architecture & the MontiArc Tool

Distributed interactive systems communicate via messages on a bus, discrete event signals, streams of telephone or video data, method invocation, or data structures passed between software services. We use streams, statemachines and components [BR07] as well as expressive forms of composition and refinement [PR99] for semantics. Furthermore, we built a concrete tooling infrastructure called MontiArc [HRR12] for architecture design and extensions for states [RRW13b]. MontiArc was extended to describe variability [HRR+11] using deltas [HRRS11] and evolution on deltas [HRRS12]. [GHK+07] and [GHK+08] close the gap between the requirements and the logical architecture and [GKPR08] extends it to model variants. Co-evolution of architecture is discussed in [MMR10] and a modeling technique to describe dynamic architectures is shown in [HRR98].

## Compositionality & Modularity of Models

[HKR+09] motivates the basic mechanisms for modularity and compositionality for modeling. The mechanisms for distributed systems are shown in [BR07] and algebraical-

ly underpinned in [HKR+07]. Semantic and methodical aspects of model composition [KRV08] led to the language workbench MontiCore [KRV10] that can even develop modeling tools in a compositional form. A set of DSL design guidelines incorporates reuse through this form of composition [KKP+09]. [Völ11] examines the composition of context conditions respectively the underlying infrastructure of the symbol table. Modular editor generation is discussed in [KRV07a].

## Semantics of Modeling Languages

The meaning of semantics and its principles like underspecification, language precision and detailedness is discussed in [HR04]. We defined a semantic domain called "System Model" by using mathematical theory. [RKB95, BHP+98] and [GKR96, KRB96]. An extended version especially suited for the UML is given in [BCGR09b] and in [BCGR09a] its rationale is discussed. [BCR07a, BCR07b] contain detailed versions that are applied on class diagrams in [CGR08]. [MRR11a, MRR11b] encode a part of the semantics to handle semantic differences of activity diagrams and [MRR11e] compares class and object diagrams with regard to their semantics. In [BR07], a simplified mathematical model for distributed systems based on black-box behaviors of components is defined. Metamodeling semantics is discussed in [EFLR99]. [BGH+97] discusses potential modeling languages for the description of an exemplary object interaction, today called sequence diagram. [BGH+98] discusses the relationships between a system, a view and a complete model in the context of the UML. [GR11] and [CGR09] discuss general requirements for a framework to describe semantic and syntactic variations of a modeling language. We apply these on class and object diagrams in [MRR11e] as well as activity diagrams in [GRR10]. [Rum12] embodies the semantics in a variety of code and test case generation, refactoring and evolution techniques. [LRSS10] discusses evolution and related issues in greater detail.

## Evolution & Transformation of Models

Models are the central artifact in model driven development, but as code they are not initially correct and need to be changed, evolved and maintained over time. Model transformation is therefore essential to effectively deal with models. Many concrete model transformation problems are discussed: evolution [LRSS10, MMR10, Rum04], refinement [PR99, KPR97, PR94], refactoring [Rum12, PR03], translating models from one language into another [MRR11c, Rum12] and systematic model transformation language development [Wei12]. [Rum04] describes how comprehensible sets of such transformations support software development, maintenance and [LRSS10] technologies for evolving models within a language and across languages and linking architecture descriptions to their implementation [MMR10]. Automaton refinement is discussed in [PR94, KPR97], refining pipe-and-filter architectures is explained in [PR99]. Refactorings of models are important for model driven engineering as discussed in [PR03, Rum12]. Translation between languages, e.g., from class diagrams into Alloy [MRR11c] allows for comparing class diagrams on a semantic level.

## Variability & Software Product Lines (SPL)

Many products exist in various variants, for example cars or mobile phones, where one manufacturer develops several products with many similarities but also many variations. Variants are managed in a Software Product Line (SPL) that captures the commonalities as well as the differences. Feature diagrams describe variability in a top down fashion, e.g., in the automotive domain [GHK+08] using 150% models. Reducing overhead and associated costs is discussed in [GRJA12]. Delta modeling is a bottom up technique starting with a small, but complete base variant. Features are added (that sometimes also modify the core). A set of applicable deltas configures a system variant. We discuss the application of this technique to Delta-MontiArc [HRR+11, HRR+11] and to Delta-Simulink [HKM+13]. Deltas can not only describe spacial variability but also temporal variability which allows for using them for software product line evolution [HRRS12]. [HHK+13] describes an approach to systematically derive delta languages. We also apply variability to modeling languages in order to describe syntactic and semantic variation points, e.g., in UML for frameworks [PFR02]. And we specified a systematic way to define variants of modeling languages [CGR09] and applied this as a semantic language refinement on Statecharts in [GR11].

## Cyber-Physical Systems (CPS)

Cyber-Physical Systems (CPS) [KRS12] are complex, distributed systems which control physical entities. Contributions for individual aspects range from requirements [GRJA12], complete product lines [HRRW12], the improvement of engineering for distributed automotive systems [HRR12] and autonomous driving [BR12a] to processes and tools to improve the development as well as the product itself [BBR07]. In the aviation domain, a modeling language for uncertainty and safety events was developed, which is of interest for the European airspace [ZPK+11]. A component and connector architecture description language suitable for the specific challenges in robotics is discussed in [RRW13b]. Monitoring for smart and energy efficient buildings is developed as Energy Navigator toolset [KPR12, FPPR12, KLPR12].

## State Based Modeling (Automata)

Today, many computer science theories are based on state machines in various forms including Petri nets or temporal logics. Software engineering is particularly interested in using state machines for modeling systems. Our contributions to state based modeling can currently be split into three parts: (1) understanding how to model object-oriented and distributed software using statemachines resp. Statecharts [GKR96, BCR07b, BCGR09b, BCGR09a], (2) understanding the refinement [PR94, RK96, Rum96] and composition [GR95] of statemachines, and (3) applying statemachines for modeling systems. In [Rum96] constructive transformation rules for refining automata behavior are given and proven correct. This theory is applied to features in [KPR97]. Statemachines are embedded in the composition and behavioral specifications concepts of Focus [BR07]. We apply these techniques, e.g., in MontiArcAutomaton [RRW13a] as well as in building management systems [FLP+11].

## Robotics

Robotics can be considered a special field within Cyber-Physical Systems which is defined by an inherent heterogeneity of involved domains, relevant platforms, and challenges. The engineering of robotics applications requires composition and interaction of diverse distributed software modules. This usually leads to complex monolithic software solutions hardly reusable, maintainable, and comprehensible, which hampers broad propagation of robotics applications. The MontiArcAutomaton language [RRW13a] extends ADL MontiArc and integrates various implemented behavior modeling languages using MontiCore [RRW13b] that perfectly fits Robotic architectural modelling. The LightRocks [THR+13] framework allows robotics experts and laymen to model robotic assembly tasks.

## Automotive, Autonomic Driving & Intelligent Driver Assistance

Introducing and connecting sophisticated driver assistance, infotainment and communication systems as well as advanced active and passive safety-systems result in complex embedded systems. As these feature-driven subsystems may be arbitrarily combined by the customer, a huge amount of distinct variants needs to be managed, developed and tested. A consistent requirements management that connects requirements with features in all phases of the development for the automotive domain is described in [GRJA12]. The conceptual gap between requirements and the logical architecture of a car is closed in [GHK+07, GHK+08]. [HKM+13] describes a tool for delta modeling for Simulink [HKM+13]. [HRRW12] discusses means to extract a well-defined Software Product Line from a set of copy and paste variants. Quality assurance, especially of safety-related functions, is a highly important task. In the Carolo project [BR12a, BR12b], we developed a rigorous test infrastructure for intelligent, sensor-based functions through fully-automatic simulation [BBR07]. This technique allows a dramatic speedup in development and evolution of autonomous car functionality, and thus, enables us to develop software in an agile way [BR12a]. [MMR10] gives an overview of the current state-of-the-art in development and evolution on a more general level by considering any kind of critical system that relies on architectural descriptions. As tooling infrastructure, the SSElab storage, versioning and management services [HKR12] are essential for many projects.

## Energy Management

In the past years, it became more and more evident that saving energy and reducing $CO_2$ emissions is an important challenge. Thus, energy management in buildings as well as in neighbourhoods becomes equally important to efficiently use the generated energy. Within several research projects, we developed methodologies and solutions for integrating heterogeneous systems at different scales. During the design phase, the Energy Navigators

Active Functional Specification (AFS) [FPPR12, KPR12] is used for technical specification of building services already. We adapted the well-known concept of statemachines to be able to describe different states of a facility and to validate it against the monitored values [FLP⁺11]. We show how our data model, the constraint rules and the evaluation approach to compare sensor data can be applied [KLPR12].

## Cloud Computing & Enterprise Information Systems

The paradigm of Cloud Computing is arising out of a convergence of existing technologies for web-based application and service architectures with high complexity, criticality and new application domains. It promises to enable new business models, to lower the barrier for web-based innovations and to increase the efficiency and cost-effectiveness of web development. Application classes like Cyber-Physical Systems [KRS12], Big Data, App and Service Ecosystems bring attention to aspects like responsiveness, privacy and open platforms. Regardless of the application domain, developers of such systems are in need for robust methods and efficient, easy-to-use languages and tools. We tackle these challenges by perusing a model-based, generative approach [PR13]. The core of this approach are different modeling languages that describe different aspects of a cloud-based system in a concise and technology-agnostic way. Software architecture and infrastructure models describe the system and its physical distribution on a large scale. We apply cloud technology for the services we develop, e.g., the SSELab [HKR12] and the Energy Navigator [FPPR12, KPR12] but also for our tool demonstrators and our own development platforms. New services, e.g.,c collecting data from temperature, cars etc. are easily developed.